\documentclass[twocolumn]{autart}    

\usepackage{graphicx}          
\usepackage[dvips]{epsfig}    
                               
\usepackage{subfigure}
\usepackage{algorithm}
\usepackage{algorithmic}
\usepackage{amsmath}
\usepackage{mathbbol}
\usepackage{bm}
\usepackage{amsfonts} 
\usepackage{breqn}
\usepackage{xr}
\usepackage{xcolor}

\usepackage{ragged2e}
\usepackage{mathrsfs}

\usepackage{bbold}

\makeatletter
\def\amsbb{\use@mathgroup \M@U \symAMSb}
\makeatother

\newcommand{\specnorm}[1]{{\left\vert\kern-0.25ex\left\vert\kern-0.25ex\left\vert #1 
    \right\vert\kern-0.25ex\right\vert\kern-0.25ex\right\vert}}

\newcommand{\WideLaplacian}{\boldsymbol{\mathcal{L}}}
\newcommand{\WideMAtrixA}{\boldsymbol{\mathcal{A}}}

\newcommand{\Lp}{\boldsymbol{\mathcal{L}}}
\newcommand{\Li}{\boldsymbol{\mathcal{L}}}
\newcommand{\Lc}{\boldsymbol{\mathcal{L}}}
\newcommand{\Lh}{\boldsymbol{\mathcal{H}}}

\newcommand{\Vone}{\mathbb{1}}
\newcommand{\Vzero}{\mathbb{0}}

\begin{document}

\begin{frontmatter}

\title{Multiplex PI Control for Consensus in Networks of Heterogeneous Linear Agents} 

\thanks[footnoteinfo]{This paper was not presented at any IFAC 
meeting. Corresponding author Tel. (+39)0817683854.}

\author[Da]{Daniel Burbano}\ead{danielalberto.burbanolombana@unina.it},    
\author[Da,Ma]{Mario di Bernardo}\ead{mario.dibernardo@unina.it},               

\address[Da]{Department of Electrical Engineering and Information Technology\\University of Naples Federico II,Via Claudio 21, 80125 Naples, Italy}  
\address[Ma]{Department of Engineering Mathematics, University of Bristol, U.K.}             


\begin{abstract}                          
In this paper, we propose a multiplex proportional-integral approach, for solving consensus problems in networks of heterogeneous nodes dynamics affected by constant disturbances. The proportional and integral actions are deployed on two different layers across the network, each with its own topology. Sufficient conditions for convergence are derived that depend upon the structure of the network, the parameters characterizing the control layers and the node dynamics. The effectiveness of the theoretical results is illustrated using a power network model as a representative example.
\end{abstract}

\end{frontmatter}

\section{Introduction}
Steering the collective behaviour of a network of dynamical agents towards a desired common target state is a fundamental problem in network control \cite{CHEN2013,Cornelius2013,LiuS2011}.  A paradigmatic example is the problem of achieving consensus, where the goal is for all agent states in the network to asymptotically converge towards each other \cite{OlfatFM2007}. The existing literature on consensus is vast and many extensions and different approaches have been proposed, e.g. \cite{Ren2007c,ReCa:11}. Often, it is assumed that the agent dynamics are either trivial (simple or higher order integrators \cite{Ren2007HO}) or identical across the network \cite{ScardS2009,Li2010}. Also, the presence of disturbances and noise is often neglected.


In contrast, many real world applications are modelled as networks of heterogeneous dynamical systems, and are affected by disturbances and noise. Take for instance a network of power generators, as those considered in \cite{Hill2006,Motter2013power}. Different power sources and transmission lines, multiple load variations, and even communication failures between generators make the network highly heterogeneous.
\begin{figure}[tbp]
\centering {
\subfigure[]
{\label{fig:introDMPIa}
{\includegraphics[scale=0.24]{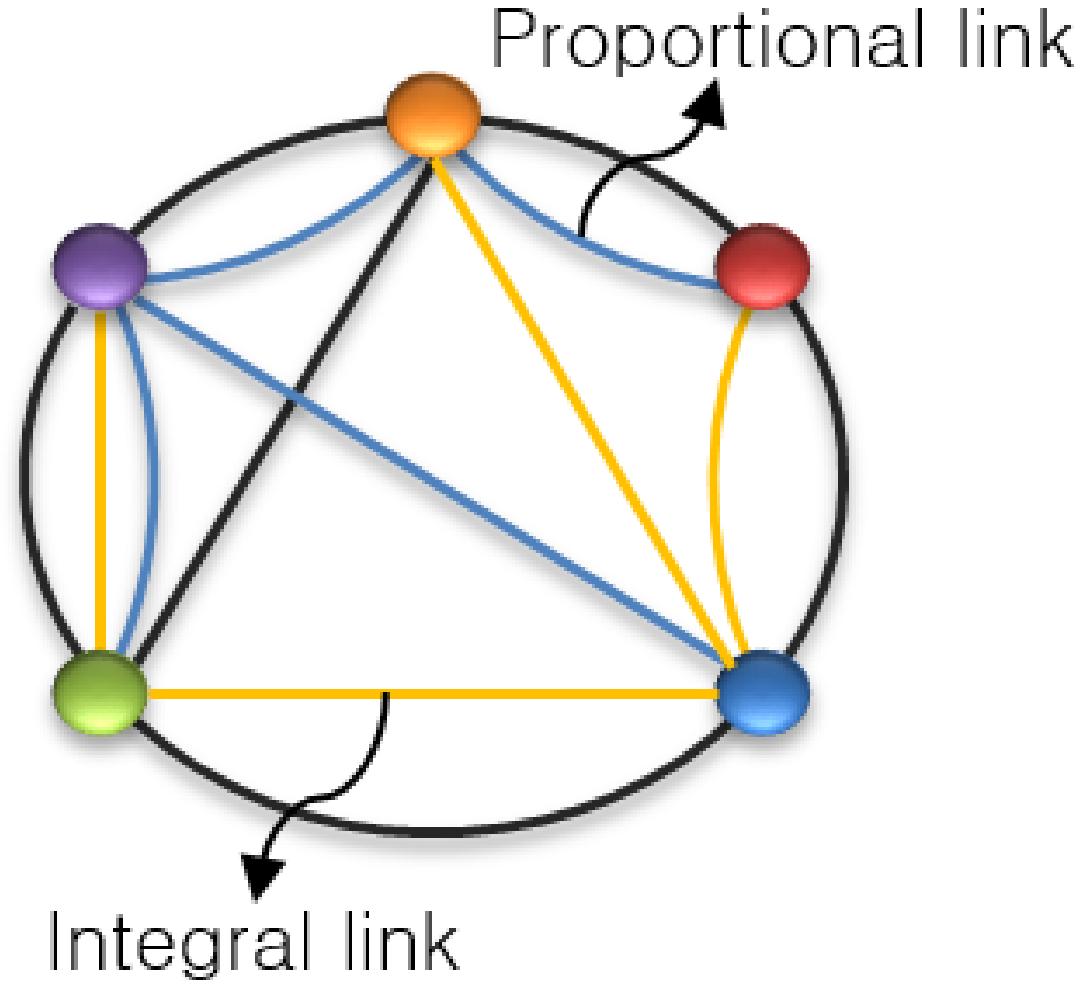}}}
\subfigure[]
{\label{fig:introDMPIb}
{\includegraphics[scale=0.19]{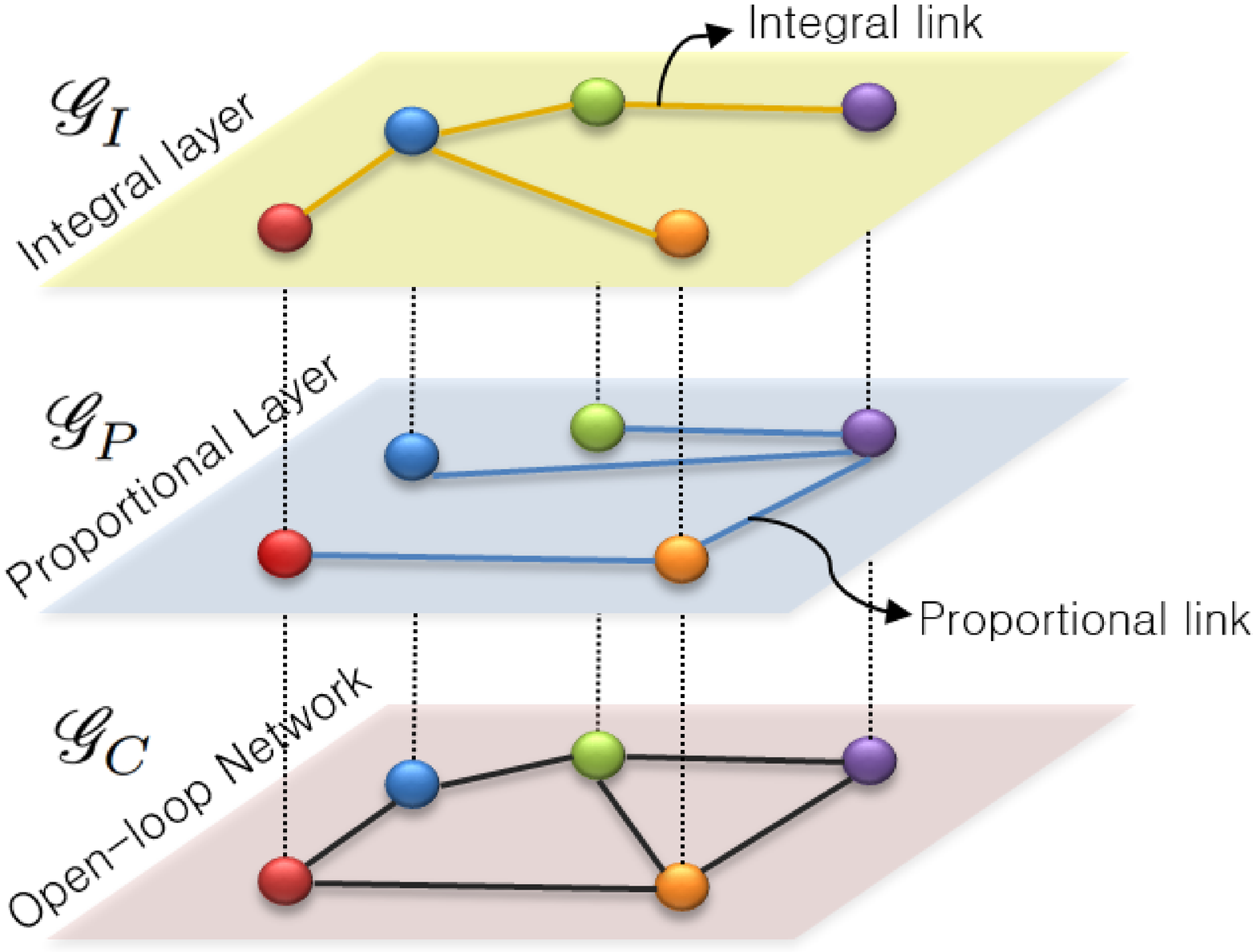}}}
}
\caption{(a): The network to be controlled is represented by black links and the blue and yellow connections represent the additional proportional and integral links that are used for control. (b) Multiplex representation of a network controlled by proportional and integral distributed controllers.}
\label{fig:introDMPI}
\end{figure}
The use of dynamic couplings implemented via the deployment of a distributed integral action has been proposed in the literature as a viable alternative to diffusive coupling when disturbances are present and/or the nodes are heterogeneous.
A distributed integral action is used, for example, in \cite{Freeman2006} to prove convergence in a  network of homogeneous first order linear systems affected by constant disturbances, while in \cite{Andreasson2014} a similar integral action is exploited to achieve consensus in homogeneous networks of simple and double integrators affected by constant disturbances. Further extensions of such distributed PI control to the case where the nodes have a more general homogeneous dynamics have been reported in \cite{Seyboth2015}. Applications have  been discussed to  achieve clock synchronization in networks of discrete-time integrators in \cite{CarliCSZ2008}, and for solving network congestion control problems in \cite{Zhang2014}. The use of distributed integral actions is also often used to achieve synchronization in power systems; see for instance \cite{Sarlette2012,Simpson-Porco2013,Andreasson2014,Bidram2014} and references therein. 
More recently, extensions have been proposed to the case where agents do not share the same dynamics. In this case the network is heterogeneous and fewer results are available particularly when the presence of disturbances, e.g. constant biases, is taken explicitly into account (see Sec. \ref{Liter_review} for a more detailed discussion of the relevant previous work in the literature).
In most of the available results, convergence is proved under the assumption that the integral action is deployed across all links in the network. Take for instance the recent work presented in \cite{Andreasson2014} or the distributed PID approach in \cite{Burbano2014ISCAS,BurbanoLombana2015} (and references therein).

In this paper we propose instead a multiplex strategy where the proportional and integral layer each possess a different structure (see Fig.~\ref{fig:introDMPIb}). The resulting closed-loop network is described by a multigraph (hypergraph) \cite{Bondy2008} which represents a class of networks recently defined as {\em multiplex networks}, which are the focus of much research attention in Physics and Applied Science (see the recent paper in Science \cite{Mucha2010}). Namely, according to the {\em multiplex PI strategy} described in this paper, two control layers are used to steer the dynamics of the open loop network offering a new degree of freedom during the design: the possibility of selecting independently the structure of the integral layer from that of the diffusive proportional one.
We show that the key analytical hurdle represented by the presence of multiple Laplacians describing each of the layers in the multigraph can be overcome so as to obtain a rigorous proof of convergence. The conditions we find are global and can be used to tune both the gains {\em and} the structure of the two control layers to achieve consensus, despite the presence of heterogeneities and constant disturbances. All the theoretical results are illustrated via representative examples that are also used to investigate the beneficial effects (in terms of stability and performance) of varying the structure of the integral layer (while keeping that of the proportional layer unchanged).


\subsection{Relevant previous work}
\label{Liter_review}
The idea of using a distributed integral action to achieve consensus in a multi-agent system has been discussed in a number of previous papers in the literature often, but not always, under the assumption of homogeneous node dynamics. Here we give a brief overview of some key previous work to better expound our results in the context of the existing literature. We wish to emphasise that the use of distributed integral actions is also common practice to achieve synchronization and frequency control in power grids, see for example \cite{Sarlette2012,Simpson-Porco2013} and references therein. In \cite{Freeman2006}, a distributed PI protocol is presented to achieve consensus in a multi-agent system. The proof of convergence is obtained for a network of scalar homogeneous agents with possibly different gains for the P and I actions, but such that they are either both present on an edge or not. Basically, while the strength of the P and I couplings can be modulated independently, the structure of the P and I interconnections is assumed to be the same. Note that this assumption is crucial for the proof of convergence presented therein as is the hypothesis that all nodes share the same dynamics. 
This is also the case for the work presented in \cite{Andreasson2014} where a distributed integral action is deployed to achieve consensus in a network of scalar, homogeneous agents in the presence of constant disturbances. 
The idea of using integrators on the Laplacian dynamics for arbitrary homogeneous linear systems is also discussed in \cite{ScardS2009}.

A more general approach is presented in the seminal work \cite{Wieland2009,Wieland2011} where the problem is considered of achieving output consensus in a network of  heterogeneous linear systems, subject to arbitrary (non-constant) disturbances.  Therein, the internal model principle  is used to prove that exact (non-trivial) output synchronisation is only possible if the intersection of the agents' spectra is non-empty. In practice, agents can only synchronize to ``a trajectory generated by a dynamical system contained in the dynamics of each agent or exosystem'' (as explained in \cite{Seyboth2012}). As pointed out in \cite{Seyboth2012} this condition is not always satisfied, as for example, in a network of heterogeneous harmonic oscillators. Also the structure of the proportional and integral layers is assumed to be the same. 
The use of the internal model principle is also adopted in \cite{Lunze2012} to study synchronization of heterogeneous agents.
The internal model principle is further exploited in \cite{Bai2010} to extend the previous work in \cite{Freeman2006} and prove convergence in the presence of time-varying inputs including polynomial inputs of known order and sinusoidal inputs with known frequencies. It is also used in \cite{Bai2014} together with incremental passivity to prove convergence in a network of nonlinear systems under a certain class of disturbances. In particular, it is shown that consensus is achieved if the Laplacian describing the integral layer is symmetric. Also, the integral action is based on the output of an internal model system and the disturbance is assumed to be generated by a known dynamical model.
Finally, synchronization of heterogeneous nonlinear systems is studied in a number of papers in the literature as for example in  \cite{ZhongOct,DeLellis2015} and extensions of the internal model principle to this class of systems has been recently presented in \cite{Burger2014,Wieland2013}.
When compared to the existing literature, in this paper we present a different approach based on the deployment of a distributed PI action in networks of heterogeneous linear agents in the presence of constant disturbances (or affine terms) and, unlike other previous work, when the control layers have different structures. We wish to emphasize that arguments based on the internal model principle (such as those reported in \cite{Wieland2009},\cite{Wieland2011}) to prove existence of a consensus equilibrium cannot be applied in our case (see Remark \ref{Remark:9} in Sec \ref{Sec:conv:equ} for further details). 
%
%
\section{Preliminaries}
\label{sec:PR}
We denote by $\mathbf{I}_N$ the identity matrix of dimension $N\times N$; by $\Vzero_{M\times N}$ a matrix of zeros of dimension $M\times N$, and by $\Vone_N$  a $N\times 1$ vector with unitary elements. The Frobenius norm is denoted by $\left\| \cdot \right\|$ while the spectral norm by $\specnorm{\cdot}$. A diagonal matrix, say $\mathbf{D}$, with diagonal elements $d_1, \ldots,d_N$ is indicated by $\mathbf{D}=\mbox{diag}\{d_1,\ldots,d_N\}$. The determinant of a matrix is denoted by $\det(.)$, $\lambda _k(\mathbf{A})$ denotes the $k$-th eigenvalue of a squared matrix $\mathbf{A}$, and $\mathbf{A}'=\mathbf{A}+\mathbf{A}^T$ denotes the symmetric part of a matrix. 
\begin{prop}
Given two vectors $\mathbf{v}_1\in \amsbb{R}^{n\times 1}$, $\mathbf{v}_2\in \amsbb{R}^{m\times 1}$ and two matrices $\mathbf{Q}_1\in \amsbb{R}^{m\times n}$, $\mathbf{Q}_2\in \amsbb{R}^{m\times m}$, some algebraic manipulations yield
\begin{equation}
\label{pos:eq:1}
2\mathbf{v}_1^T\mathbf{Q}_1^T\mathbf{Q}_2\mathbf{v}_2 \le \varepsilon\mathbf{v}_1^T\mathbf{Q}_1^T\mathbf{Q}_1\mathbf{v}_1+\frac{1}{\varepsilon}\mathbf{v}_2^T\mathbf{Q}_2^T\mathbf{Q}_2\mathbf{v}_2,\forall \varepsilon>0 
\end{equation}
\end{prop}
\begin{pf*}{Proof.}
Consider the $m \times 1$ vector $ {a{\mathbf{Q}_1}{\mathbf{v}_1} \pm b{\mathbf{Q}_2}{\mathbf{v}_2}} $ with $a,b \in \amsbb{R}^{+}$. From its quadratic form one has $\left( {a{\mathbf{Q}_1}{\mathbf{v}_1} \pm b{\mathbf{Q}_2}{\mathbf{v}_2}} \right)^T\left( {a{\mathbf{Q}_1}{\mathbf{v}_1} \pm b{\mathbf{Q}_2}{\mathbf{v}_2}} \right)\ge 0$
and
\[{a^2}\mathbf{v}_1^T\mathbf{Q}_1^T{\mathbf{Q}_1}{\mathbf{v}_1} \pm 2ab\mathbf{v}_1^T\mathbf{Q}_1^T{\mathbf{Q}_2}{\mathbf{v}_2} + {b^2}\mathbf{v} _2^T\mathbf{Q}_2^T{\mathbf{Q}_2}{\mathbf{v}_2} \ge 0\]
then, dividing both sides of the inequality by $ab$ we have that $2\mathbf{v}_1^T\mathbf{Q}_1^T{\mathbf{Q}_2}{\mathbf{v}_2} \le {a}/{b}\mathbf{v}_1^T\mathbf{Q}_1^T{\mathbf{Q}_1}{\mathbf{v}_1} + {b}/{a}\mathbf{v}_2^T\mathbf{Q}_2^T{\mathbf{Q}_2}{\mathbf{v}_2}$. Finally, setting $\varepsilon=a/b$ we obtain \eqref{pos:eq:1}.
\end{pf*}
\begin{lem}
\label{lemm:quadratic_form}
Given a symmetric matrix $\mathbf{A}\in \amsbb{R}^{n\times n}$, denoting by ${\lambda _{\min }}(\mathbf{A})$ and  ${\lambda _{\max }}(\mathbf{A})$ the smallest and largest eigenvalues of $\mathbf{A}$, the following statements are true \cite{RAH_CRJ_1987}
\begin{eqnarray}
\label{symm:bound}
{\lambda _{\min }}(\mathbf{A}) {{ {{\mathbf{v}}} }^T} {{\mathbf{v}}} \le {{ {{\mathbf{v}}} }^T}\mathbf{A} {{\mathbf{v}}} \le {\lambda _{\max }}(\mathbf{A}){{ {{\mathbf{v}}} }^T} {{\mathbf{v}}},\forall \mathbf{v} \in \amsbb{R}^{n\times 1}\\
\label{specnorm:bou}
\specnorm{ \mathbf{A} } = \mathop {\max }\limits_k \left\{ {\left| {{\lambda _k}(\mathbf{A})} \right|} \right\}\leq{\left\| \mathbf{A} \right\|}\\
\label{eig_bloksim}
{\lambda _{\min }}(\mathbf{A}) \le {\lambda _{\min }}({\mathbf{A}_o}) \le {\lambda _{\max }}({\mathbf{A}_o}) \le {\lambda _{\max }}(\mathbf{A})
\end{eqnarray}
where ${\mathbf{A}_o}\in \amsbb{R}^{k\times k}$ is a principal sub-matrix of $\mathbf{A}$ {(See Corollary 8.4.6 in \cite{Bernstein2009})}.
\end{lem}
\begin{lem}\cite{Bernstein2009}
\label{kron:rel}
Given the matrices $\mathbf{A}$,$\mathbf{B}$, $\mathbf{C}$ and $\mathbf{D}$ of appropriate dimensions, the Kronecker product satisfies the following properties
\begin{eqnarray}
\label{kron:rel:a}
(\mathbf{A} \otimes \mathbf{B}) + (\mathbf{A} \otimes \mathbf{C) = \mathbf{A} \otimes (\mathbf{B} + \mathbf{C})}   \\
\label{kron:rel:b}
(\mathbf{A} \otimes \mathbf{B})(\mathbf{A} \otimes \mathbf{D}) = \mathbf{AB} \otimes \mathbf{BD}  \\
\label{kron:rel:c}
\specnorm{(\mathbf{A} \otimes \mathbf{B})} = \specnorm{\mathbf{A}}\specnorm{\mathbf{B}} 
\end{eqnarray}
\end{lem}
\subsection{Algebraic graph theory}
 An {\em undirected graph} $\mathscr{G}$ is a pair defined by $\mathscr{G} = \left( {\mathcal{N},\mathcal{E}} \right)$ where $\mathcal{N} = \left\{ {{1},{2}, \cdots ,{N}} \right\}$ is the finite set of $N$ node indices; $\mathcal{E} \subset \mathcal{N} \times \mathcal{N}$ is the set containing the $P$ edges between the nodes. We assume each edge has an associated weight denoted by $w_{ij} \in \amsbb{R}^{+}$ for all $i,j \in \mathcal{N}$. The weighed \textit{adjacency matrix} $\WideMAtrixA(\mathscr{G})\in {\amsbb{R}^{N \times N}}$ with $\mathcal{A}_{ij}$ entries, is defined as $\mathcal{A}_{ij}(\mathscr{G})=w_{ij}$ if there is an edge from node $i$ to node $j$ and zero otherwise.
Similarly, the \textit{Laplacian matrix} $\WideLaplacian(\mathscr{G})\in {\amsbb{R}^{N \times N}}$ is defined as the matrix whose elements ${\mathcal{L}_{ij}}(\mathscr{G}) = \sum\nolimits_{j = 1,j \ne i}^N {{w_{ij}}} $ if $i=j$ and $-{{w_{ij}}}$ otherwise. Thus, the Laplacian matrix can be recast in compact form as $\WideLaplacian(\mathscr{G}) = \mbox{diag}\{\WideMAtrixA(\mathscr{G})\Vone_N\} - \WideMAtrixA(\mathscr{G})$, where the matrix $\mbox{diag}\{\WideMAtrixA(\mathscr{G})\Vone_N\}$ is often called the degree matrix of the graph $\mathscr{G}$. Given two graphs sharing the same set of nodes $\mathscr{G}_1=(\mathcal{N},\mathcal{E}_{1})$ and $\mathscr{G}_2=(\mathcal{N},\mathcal{E}_{2})$, we define the {\em projection graph} as the graph $\mbox{proj}(\mathscr{G}_1,\mathscr{G}_2):=(\mathcal{N},\mathcal{E}_{p})$ with associate adjacency matrix $\WideMAtrixA_p:=\WideMAtrixA(\mathscr{G}_1) + \WideMAtrixA(\mathscr{G}_2)$.
\begin{defn}\cite{Lu2006214}
We say that an $N\times N$ matrix $\boldsymbol{\mathcal{S}}  = [{\mathcal{S}_{ij}}], \forall i,j\in \mathcal{N}$  belongs to the set $\amsbb{W}$ if it verifies the following properties:
\begin{enumerate}
	\item ${{\mathcal{S}}_{ij}} \le 0,\,i \ne j,$ and ${{\mathcal{S}}_{ii}} =  - \sum\limits_{j = 1,j \ne i}^N {{{\mathcal{S}}_{ij}}}$,
	\item its eigenvalues in ascending order are such that $\lambda _1(\boldsymbol{\mathcal{S}})=0$ while all the others, $\lambda _k(\boldsymbol{\mathcal{S}})$, $k\in \{2,\cdots,N\}$, are real and positive.
\end{enumerate}
\label{def_laplacian}
\end{defn}
The set of matrices $\amsbb{W}$ defined above are in fact a special instance of $M$-matrices as defined in  \cite{Poole1974}.
Note that the Laplacian matrix $\WideLaplacian$ belongs to the set $\amsbb{W}$ if its associated graph $\mathscr{G}$ is connected \cite{OlfatFM2007}.
Next, we present a decomposition of the Laplacian matrix that will be crucial for the derivations reported in the rest of the paper. As suggested in \cite{BurbanoLombana2015} such a decomposition is particularly useful to prove convergence in the presence of heterogeneous nodes.
\begin{lem}\cite{BurbanoLombana2015}
\label{lemm:simmetric_L}
Let $\WideLaplacian\in\amsbb{W}$ be the Laplacian matrix of an undirected and connected graph $\mathscr{G}$, then $\WideLaplacian$ can be written in block form as $\WideLaplacian=\mathbf{R} \mathbf{\Lambda} \mathbf{R}^{-1}$, where $\mathbf{R}$ is an orthonormal matrix defined with its inverse as
\begin{equation}
\label{block:decompo}
\begin{array}{l}
\mathbf{R} =  \left[ {\begin{array}{*{20}{c}}
   1 & {N{\mathbf{R}_{21}^T}}  \\
   {{\Vone_{N - 1}}} & {N{\mathbf{R}_{22}^T}}  \\
\end{array}} \right],\,{\mathbf{R}^{-1}} = \left[ {\begin{array}{*{20}{c}}
   {{r_{11}}} & {{\mathbf{R}_{12}}}  \\
   {{\mathbf{R}_{21}}} & {{\mathbf{R}_{22}}}  \\
\end{array}} \right]
\end{array}
\end{equation}
with \begin{eqnarray}
r_{11} = \frac{1}{N}, \qquad \mathbf{R}_{12}= \frac{1}{N}\Vone_{N - 1}^T,\label{eq:blockdef}
\end{eqnarray}
$\mathbf{R}_{21}\in {\amsbb{R}^{{(N - 1) \times 1} }}$, $\mathbf{R}_{22}\in {\amsbb{R}^{{(N - 1) \times (N - 1)}}}$ being blocks of appropriate dimensions, $\mathbf{\Lambda}  = \mbox{diag}\left\{ {0,{\lambda _2(\WideLaplacian)}, \cdots ,{\lambda _N(\WideLaplacian)}} \right\}$ with $0 = {\lambda _1(\WideLaplacian)} < {\lambda _2(\WideLaplacian)} \le  \cdots  \le {\lambda _N(\WideLaplacian)}$ being the eigenvalues of $\WideLaplacian$ in ascending order. Also, the blocks in $\mathbf{R}$ and $\mathbf{R}^{-1}$ must fulfill the following conditions
\begin{eqnarray}
\label{prop:U:1}
{r_{11}}{\mathbf{I}_n} + ({\mathbf{R}_{12}}{\Vone_{N - 1}} \otimes {\mathbf{I}_n}) = {\mathbf{I}_n}\\
\label{prop:U:2}
({\mathbf{R}_{21}} \otimes {\mathbf{I}_n}) + ({\mathbf{R}_{22}}{\Vone_{N - 1}} \otimes {\mathbf{I}_n}) = {\Vzero_{(n(N - 1) \times 1)}}\\
\label{prop:U:3}
({\mathbf{R}_{21}}\mathbf{R}_{21}^T \otimes {\mathbf{I}_n}) + ({\mathbf{R}_{22}}\mathbf{R}_{22}^T \otimes {\mathbf{I}_n}) = \frac{1}{N}({\mathbf{I}_{N - 1}} \otimes {\mathbf{I}_n})\\
\label{prop:U:4}
{r_{11}}(\mathbf{R}_{21}^T \otimes {\mathbf{I}_n}) + ({\mathbf{R}_{12}}\mathbf{R}_{22}^T \otimes {\mathbf{I}_n}) = {\Vzero_{(1 \times n(N - 1))}}\\
\label{prop:U:5}
({\mathbf{R}_{21}}\mathbf{R}_{21}^T \otimes {\mathbf{I}_n}) =  ({\mathbf{R}_{22}}{\Vone_{N-1}\Vone_{N-1}^T}\mathbf{R}_{22}^T \otimes {\mathbf{I}_n})\\
\label{prop:U:5norm}
\specnorm{ {({\mathbf{R}_{22}} \otimes {\mathbf{I}_n})} } \le \frac{1}{{\sqrt N }}\\
\label{bound_R21_termsofN}
\left\| \mathbf{R}_{21}\right\|\leq\sqrt{N-1}\specnorm{\mathbf{R}_{22}} \leq \sqrt{(N-1)/N}\\
\label{transpose_R}
{\mathbf{R}^T=N\mathbf{R}^{-1}}\\
\label{prop:U:7inv}
N\mathbf{R}_{22}^T = (\mathbf{I}_{N-1}+\Vone_{N-1}\Vone_{N-1}^T)^{-1}\mathbf{R}_{22}^{-1}
\end{eqnarray}
\end{lem}
\begin{pf*}{Proof.}
See appendix \ref{Appendix_II}.
\end{pf*}

%
\begin{defn} A multigraph, is the set of M graphs $\mathscr{M}:= \{\mathscr{G}_1, \cdots, \mathscr{G}_M  \}$ called layers of $\mathscr{M}$, where all the graphs in $\mathscr{M}$ share the same set of nodes, that is $\mathscr{G}_k=\left( {\mathcal{N},\mathcal{E}_k} \right)$, for $k \in \{1,\cdots,M \}$.
\end{defn}
%
%
%
%
\section{Problem statement and multiplex PI control}
We consider the problem of achieving consensus in a network of $N$ agents governed by open-loop heterogeneous dynamics of the form
\begin{equation}
\label{eq:sys:1}
{{\dot {\mathbf{x}}}_i}(t) = \mathbf{A}_i{\mathbf{x}_i}(t) + {\mathbf{b}_i} - \sigma \sum\nolimits_{j = 1}^N {{\mathcal{L}_{C,ij}}{\mathbf{x}_j(t)}} +{\mathbf{u}_i}(t)	
\end{equation}
for all $i \in \mathcal{N}$, where $\mathbf{x}_i(t)\in \amsbb{R}^{n\times 1}$ represents the state of the $i$-th agent, $\mathbf{A}_i\in \amsbb{R}^{n\times n}$ is the intrinsic node dynamic matrix,  ${\mathbf{b}_i}\in \amsbb{R}^{n\times 1}$ is some constant bias (or constant disturbance) acting on each node, $\sigma$ is a non-negative constant modelling the global coupling strength among any pair of nodes, ${\mathcal{L}_{C,ij}}$ are the elements of  the Laplacian matrix $\Lc_C$ of the weighed graph $\mathscr{G}_{C}:=(\mathcal{N},\mathcal{E}_{C})$ representing the open-loop network to be controlled (see Fig. \ref{fig:introDMPIb}), and $\mathbf{u}_i(t)\in \amsbb{R}^{n\times 1}$ is the control input. 
In this paper we assume that at least one bias $\mathbf{b}_i \ne \Vzero_{(n\times 1)}$ for some $i \in \mathcal{N}$. In so doing, the trivial solution is excluded  that is associated to the case where all the agent dynamics $\mathbf{A}_i$ are exponentially stable with null biases. Indeed, in this case all nodes would achieve consensus onto zero and no distributed control action would be required. 
\noindent 
\begin{defn} 
\label{def:1}
Network \eqref{eq:sys:1} is said to achieve admissible consensus if, for any set of initial conditions $x_i(0)=x_{i0}$, there exists some non negative constant $W$ such that $\mathop {\lim }\nolimits_{t \to \infty } {\left\| {\mathbf{x}_j(t)-\mathbf{x}_i(t) } \right\|=0}$ for $\ i,j\in \mathcal{N}$ and $\left\| {\mathbf{u}_i(t)} \right\|  <  W <+ \infty$, for all $t\geq 0$.
\end{defn}

%
%
The problem we shall solve is to find bounded and distributed control inputs $\mathbf{u}_i(t)$, such that all states $\mathbf{x}_i(t)$ converge asymptotically towards each other, i.e. admissible consensus. We then propose the use of a distributed multiplex PI control strategy, obtained by setting:
\begin{equation}
\label{eq:cont:2b}
\begin{split}
{\mathbf{u}_i}(t) &=   \sigma_{P} \sum\limits_{j = 1}^N {{\alpha _{ij}}{(\mathbf{x}_j(t)-\mathbf{x}_i(t))}}\\ 
& \quad + \sigma_{I} \sum\limits_{j = 1}^N {{\beta _{ij}}\int\limits_0^t {{(\mathbf{x}_j(\tau)-\mathbf{x}_i(\tau))}d\tau } } 
\end{split}
\end{equation}
where the non-negative constants $\alpha_{ij}\ge 0 $ and $\beta_{ij}\ge 0$ represent the control strengths of the proportional and integral control actions respectively (we do not consider self-loops, that is $\alpha_{ii}=\beta_{ii}=0$). It is important to highlight that this controller allows the deployment of proportional and integral actions independently from each other ($\alpha_{ij}=0$ or $\beta_{ij}=0$ for some $i$,$j\in \mathcal{N}$, $i \ne j$). The constants $\sigma_P$, $\sigma_I \in \amsbb{R}^+$ are additional parameters modulating globally the contribution of each control layer with respect to each other. 

Equation \eqref{eq:cont:2b} effectively defines two control layers each represented by a different weighted graph $\mathscr{G}_{P}:=(\mathcal{N},\mathcal{E}_{P})$ for the proportional layer and $\mathscr{G}_{I}:=(\mathcal{N},\mathcal{E}_{I})$ for the integral layer, where $\mathcal{E}_{P}$ is the set of edges with associated weights $\alpha_{ij}$ and $\mathcal{E}_{I}$ that with associated weights $\beta_{ij}$.
We denote the Laplacian matrices corresponding to each of these layers by
 $\Lp_P:=[\mathcal{L}_{P,ij}]$ and $\Lp_I:=[\mathcal{L}_{I,ij}]$, respectively; with their elements 
being defined as $\mathcal{L}_{P,ij} = \sum\nolimits_{j = 1,j \ne i}^N {{\alpha_{ij}}} $ and $\mathcal{L}_{I,ij} = \sum\nolimits_{j = 1,j \ne i}^N {{\beta_{ij}}} $ if $i=j$ and $\mathcal{L}_{P,ij} = -\alpha_{ij}$, $\mathcal{L}_{I,ij} = -\beta_{ij}$ otherwise.
As depicted in Fig. \ref{fig:introDMPI}, the resulting control strategy is therefore a {\em multiplex} distributed control strategy, and the closed-loop network a multiplex network associated to the multigraph $\mathscr{M}= \{\mathscr{G}_{C},\mathscr{G}_{P},\mathscr{G}_{I}  \}$.
Next, we define $\widehat{\Lc}_C:=({\Lc_C \otimes {\mathbf{I}_n}})$, $\widehat{\Lp}_P:=({\Lp_P \otimes {\mathbf{I}_n}})$, $\widehat{\Li}_I:=({\Li_I \otimes {\mathbf{I}_n}})$. Letting $\mathbf{x}(t)=[\mathbf{x}_1^T(t),\cdots, \mathbf{x}_N^T(t)]^T$ be the stack vector of all agent states and 
\begin{equation}
\label{integral:term}
{\mathbf{z}}(t) =\left[ {\mathbf{z}_1^T}(t), \ldots ,{\mathbf{z}_N^T}(t) \right]^T:= - \sigma_I {\widehat\Li}_I\int_0^t {\mathbf{x}(\tau )d\tau }
\end{equation}
the stack vector of all integral states, the overall dynamics of the closed-loop network can then be written as
\begin{equation}
\label{eq:DMPI}
\left[ {\begin{array}{*{20}{c}}
{\dot {\mathbf{x}}(t)}\\
{\dot {\mathbf{z}}(t)}
\end{array}} \right] = \left[ {\begin{array}{*{20}{c}}
{\widehat {\mathbf{A}} - \Lh}&{{\mathbf{I}_{nN}}}\\
{ - \sigma_I  \widehat{\Li}_I}&{\Vzero_{(nN\times nN)}}
\end{array}} \right]\left[ {\begin{array}{*{20}{c}}
{\mathbf{x}(t)}\\
{\mathbf{z}(t)}
\end{array}} \right] + \left[ {\begin{array}{*{20}{c}}
\mathbf{B} \\
\Vzero
\end{array}} \right]
\end{equation}
where $\widehat {\mathbf{A}}\in \amsbb{R}^{nN\times nN}$ is a block diagonal matrix encoding the node dynamics, $\widehat {\mathbf{A}}:=\mbox{diag}\left\{\mathbf{A}_1,\cdots,\mathbf{A}_N\right\}$, $\Lh:=\sigma \widehat{\Lc}_C+\sigma_P \widehat{\Lp}_P$, and $\mathbf{B}\in \amsbb{R}^{nN\times 1}$ is the stack vector of the constant biases, $\mathbf{B}:=[\mathbf{b}_1^T,\cdots,\mathbf{b}_N^T]^T$. 

Thus, the problem becomes that of finding conditions on the node dynamics, the gains $\sigma$, $\sigma_P$, and $\sigma_I$, and most importantly the structural properties of the open-loop network layer $\mathscr{G}_{C}$ and control layers $\mathscr{G}_{P}$ and $\mathscr{G}_{I}$, so as to guarantee emergence of admissible consensus in the closed-loop multiplex network \eqref{eq:DMPI}.
%
%
\section{Convergence Analysis}
\label{Sec:conv}
In this section we first show that the collective dynamics of the multiplex closed-loop network \eqref{eq:DMPI} has a unique equilibrium which is the solution of the admissible consensus problem. Then we derive some sufficient conditions guaranteeing asymptotic stability of such equilibrium.
\subsection{Consensus equilibrium}
\label{Sec:conv:equ}
\begin{prop}
\label{equil_point}
If the matrix $\mathbf{\Psi}_{11} := (1/N)\sum\nolimits_{k = 1}^N {{\mathbf{A}_k}}$ is non-singular, then the closed-loop network \eqref{eq:DMPI} has a unique equilibrium $\mathbf{x}^*:=\left( {{\Vone_N} \otimes {\mathbf{x}_\infty }} \right)$ and ${\mathbf{z}^*}: =  - ( {\widehat {\mathbf{A}}\mathbf{x}^* + \mathbf{B} } )$ where 
\begin{equation}
\label{eq:equili}
\begin{array}{l}
{\mathbf{x}_\infty }: =  - (1/N){\mathbf{\Psi}_{11}^{ - 1}}\sum\nolimits_{k = 1}^N {{\mathbf{b}_k}} 
\end{array}
\end{equation}
\end{prop}
\begin{pf*}{Proof.}
Setting the left-hand side of \eqref{eq:DMPI} to zero one has that $\mathbf{x}^* = (\Vone_N \otimes \mathbf{v})$, $\forall \mathbf{v} \in \amsbb{R}^{n\times 1}$ and ${\mathbf{z}^* } =  -\left(\widehat {\mathbf{A}}(\Vone_N \otimes \mathbf{v})  + \mathbf{B}  \right)$. From \eqref{integral:term}, we also have that $(\Vone_N^T\otimes \mathbf{I}_n)\mathbf{z}(t) = \Vzero_{nN\times1}$, then $(\Vone_N^T\otimes \mathbf{I}_n)\mathbf{z}^*  = \Vzero_{nN\times 1}$ and we obtain 
\[
\begin{split}
(\Vone_N^T\otimes \mathbf{I}_n)\widehat {\mathbf{A}}(\Vone_N \otimes \mathbf{v}) &= -(\Vone_N^T\otimes \mathbf{I}_n)\mathbf{B}\\
(1/N)\sum\nolimits_{k = 1}^N {\mathbf{A}_k}\mathbf{v} &= -(1/N)\sum\nolimits_{k = 1}^N {{\mathbf{b}_k}} 
\end{split}
\]
then $\mathbf{v}=- (1/N){\mathbf{\Psi}_{11}^{ - 1}}\sum\nolimits_{k = 1}^N {{\mathbf{b}_k}} = \mathbf{x}_\infty$ which completes the proof.
\end{pf*}
%
%
\begin{rem}\hspace{0.2cm}
\begin{itemize}
\item Note indeed that if controller \eqref{eq:cont:2b} is able to render this equilibrium stable, it is also able to guarantee consensus of all node states $\mathbf{x}(t)$ to a constant vector $\mathbf{x}_{\infty}$ using bounded control energy. Also, the consensus trajectory can be interpreted as the solution of the ``exo-system'' given by $\dot {\mathbf{s}}(t) =\mathbf{\Psi}_{11}\mathbf{s}(t) + (1/N)\sum\nolimits_{k = 1}^N {{\mathbf{b} _k}}$. Unlike the work in \cite{Wieland2011} where the existence of the consensus equilibrium requires all the agents in the network to have eigenvalues in common; here, we just need to show that $\mathbf{\Psi}_{11}$ is a full rank matrix.

\item Note that, in the notation of \cite{Wieland2009}, our strategy corresponds to setting the matrices $\mathbf{B}_i=\mathbf{E}_i=\mathbf{C}_i=\mathbf{G}_i=\mathbf{H}_i=\mathbf{K}_i=\mathbf{I}_n$ and more importantly the matrix defining the own dynamics of the local controllers $\mathbf{F}_i=\Vzero$. Therefore, existence of the consensus equilibrium cannot be proved in our case following the arguments therein. Specifically, the assumptions of detectability made in \cite{Wieland2009} do not apply. 
\end{itemize}
\label{Remark:9}
\end{rem}
Now, to prove convergence, it suffices to guarantee that ($\mathbf{x}^*,\mathbf{z}^*$) is globally asymptotically stable. We start by shifting the origin via the state transformation $\mathbf{y}(t):=\mathbf{z}(t) + \mathbf{B}$ so that \eqref{eq:DMPI} becomes
\begin{equation}
\label{eq:PID:2}
\left[ {\begin{array}{*{20}{c}}
{\dot {\mathbf{x}}(t)}\\
{\dot {\mathbf{y}}(t)}
\end{array}} \right] = \left[ {\begin{array}{*{20}{c}}
{\widehat {\mathbf{A}} - \Lh}&{{\mathbf{I}_{nN}}}\\
{ - \sigma_I \widehat{\Li}_I}&{\Vzero_{(nN\times nN)}}
\end{array}} \right]\left[ {\begin{array}{*{20}{c}}
{\mathbf{x}(t)}\\
{\mathbf{y}(t)}
\end{array}} \right]
\end{equation}
%
%
\begin{lem} Let $\WideLaplacian_1=\mathbf{R}\mathbf{\Lambda}_1\mathbf{R}^{-1}$ and $\WideLaplacian_2=\mathbf{U}\mathbf{\Lambda}_2\mathbf{U}^{-1}$ be two generic Laplacian matrices belonging to the set $\amsbb{W}$, where $\mathbf{R}$ and $\mathbf{U}$ are block matrices with the same structure as in \eqref{block:decompo} and $\mathbf{\Lambda}_k$, $k\in\left\{1,2\right\}$ are diagonal matrices containing the eigenvalues of $\WideLaplacian_1$ and $\WideLaplacian_2$ respectively. Then,
\begin{equation}
\label{lem:dif:lap}
(\mathbf{R}^{-1}\WideLaplacian_2\mathbf{R}\otimes\mathbf{I}_n)= \left[ {\begin{array}{*{20}{c}}
{{\Vzero_{(n \times n)}}}&{{\Vzero_{(n \times (nN - 1))}}}\\
{{\Vzero_{((nN - 1) \times n)}}}&{\left( {\mathbf{T} {{\bar {\mathbf{\Lambda}} }_2}{\mathbf{T} ^T} \otimes {\mathbf{I}_n}} \right)}
\end{array}} \right]
\end{equation}
where $\mathbf{T}=N\mathbf{R}_{22}(\Vone_{N-1}\Vone_{N-1}^T+\mathbf{I}_{N-1})\mathbf{U}_{22}^T$ and $\mathbf{\bar{\Lambda}}_2=\mbox{diag}\left\{\lambda_2(\WideLaplacian_2),\cdots,\lambda_N(\WideLaplacian_2)\right\}$. Moreover, ${ {\mathbf{T} {{\bar {\mathbf{\Lambda}} }_2}{\mathbf{T} ^T}} }$ is a symmetric matrix.
\label{prop:dif_lap}
\end{lem}
\begin{pf*}{Proof.}
See Appendix \ref{Appendix_I}.
\end{pf*}
%
\subsection{Error dynamics}
Assuming that the graphs in all layers of $\mathscr{M}$ are connected, using Lemma \ref{lemm:simmetric_L} we can write $\Lc_C=\mathbf{R}\mathbf{\Lambda}_C\mathbf{R}^{-1}$,  $\Lp_P=\mathbf{U}\mathbf{\Lambda}_P\mathbf{U}^{-1}$ and $\Li_I=\mathbf{Q}\mathbf{\Lambda}_I\mathbf{Q}^{-1}$. (In Corollary 13 we relax the assumption of connectivity of the open-loop network).
%
%
%
Next we define the error dynamics given by the state transformation $\mathbf{e}(t)  = ({\mathbf{R}^{-1}}\otimes \mathbf{I}_n)\mathbf{x}(t)$; therefore, using the block representation of $\mathbf{R}^{-1}$ and letting $\bar{\mathbf{e}}(t):=[\mathbf{e}_2^T(t),\cdots,\mathbf{e}_N^T(t)]^T$ and $\bar{\mathbf{x}}(t):=[\mathbf{x}_2^T(t),\cdots,\mathbf{x}_N^T(t)]^T$, we obtain
\begin{eqnarray}
\label{eq:transformation_Ua}
 \mathbf{e}_1 (t) &=& {r_{11}}{\mathbf{x}_1}(t) + ({\mathbf{R}_{12}} \otimes {\mathbf{I}_n})\bar {\mathbf{x}}(t)\\
 \label{eq:transformation_Ub}
{{\bar {\mathbf{e}}} }(t) &=& ({\mathbf{R}_{21}} \otimes {\mathbf{I}_n}){\mathbf{x}_1}(t) + ({\mathbf{R}_{22}} \otimes {\mathbf{I}_n})\bar {\mathbf{x}}(t)
\end{eqnarray}
Thus expressing $({\mathbf{R}_{21}} \otimes {\mathbf{I}_n})$ from \eqref{prop:U:2} and substituting in \eqref{eq:transformation_Ub} yields
$${{\bar {\mathbf{e}}}} (t) = ({\mathbf{R}_{22}}\otimes \mathbf{I}_n)\left( {\bar{\mathbf{x}}(t) - (\Vone_{N-1}\otimes \mathbf{I}_n){\mathbf{x}_1}(t)  } \right)$$
note that ${{\bar {\mathbf{e}}}} (t)=\Vzero$ if and only if ${\bar{\mathbf{x}}(t) - (\Vone_{N-1}\otimes \mathbf{I}_n){\mathbf{x}_1}(t)  } =\Vzero$ since ${\mathbf{R}_{22}}$ is a full rank matrix \cite{BurbanoLombana2015}. Then, admissible consensus is achieved if $\lim_{t \to \infty }{{\bar {\mathbf{e}}}} (t) = \Vzero$ and $\left\| {\mathbf{y}(t)} \right\|\le W < +\infty, \forall t>0$.

Now, recasting \eqref{eq:PID:2} in the new coordinates $\mathbf{e}(t)$ and $\mathbf{w}(t):= {\mathbf{R}^{-1}}\mathbf{y}(t)$, and letting 
${\bar{\mathbf{\Lambda}}}_C:=\mbox{diag}\{ \lambda_2(\Lc_C),\cdots,$ $\lambda_N(\Lc_C) \}$, ${\bar{\mathbf{\Lambda}}}_P:=\mbox{diag}\{ \lambda_2(\Lp_P),\cdots,\lambda_N(\Lp_P) \}$, ${\bar{\mathbf{\Lambda}}}_I:=\mbox{diag}\{ \lambda_2(\Li_I),\cdots,\lambda_N(\Li_I) \}$
we get
\begin{equation}
 \label{eq:final:PID}
 \begin{array}{l}
{{\dot {\mathbf{e}}}}(t) = \left( {\mathbf{\Psi}  - \widehat{\Lh} } \right){\mathbf{e} }(t) + \left[ {\begin{array}{*{20}{c}}
{\Vzero_{n\times1}}\\
{\bar {\mathbf{w}}(t)}
\end{array}} \right]\\
{{\dot {\bar {\mathbf{w}}}}}(t) =  - \beta (\mathbf{T}_I {{\bar {\mathbf{\Lambda}} }_I}{\mathbf{T}_I ^T}\otimes \mathbf{I}_n){{\bar {\mathbf{e}}} }(t)
\end{array}
\end{equation}
where $\bar {\mathbf{w}}(t) := \left[ \mathbf{w}_2^T(t), \ldots , \mathbf{w}_N^T(t) \right]^T$. Note that the dynamics of $\mathbf{w}_1(t)$ can be neglected as it is trivial with null initial conditions and represents an uncontrollable and unobservable state.
The quantities in \eqref{eq:final:PID} are defined as follows
\begin{itemize}
\item $\mathbf{\Psi}$ is a block matrix defined as
\[\begin{array}{l}
\mathbf{\Psi}  := \left[ {\begin{array}{*{20}{c}}
{{\mathbf{\Psi} _{11}}}&{{\mathbf{\Psi} _{12}}}\\
{{\mathbf{\Psi} _{21}}}&{{\mathbf{\Psi} _{22}}}
\end{array}} \right]=({\mathbf{R}^{ - 1}}\otimes\mathbf{I}_n)\widehat{\mathbf{A}}({\mathbf{R}}\otimes\mathbf{I}_n)
 = \\({\mathbf{R}^{ - 1}}\otimes\mathbf{I}_n)\left[ {\begin{array}{*{20}{c}}
{\mathbf{A}_1}&{{\Vzero_{(n \times n(N - 1))}}}\\
{{\Vzero_{(n(N - 1) \times n)}}}&{\bar {\mathbf{A}}}
\end{array}} \right]({\mathbf{R}}\otimes\mathbf{I}_n)
\end{array}\]
where $\bar{\mathbf{A}} := \mbox{diag}\left\{ {\mathbf{A}_2, \cdots ,\mathbf{A}_N} \right\}$ is a block diagonal matrix.  Using properties (\ref{prop:U:1})-(\ref{prop:U:4}), we can write (see Appendix \ref{Appendix_III} for the derivation)
\begin{eqnarray}
\label{eq:PSI:11}
{\mathbf{\Psi} _{11}} &=& (1/N)\sum\nolimits_{k = 1}^N {{\mathbf{A}_k}}\\
\label{eq:PSI:12}
{\mathbf{\Psi}_{12}} &=&  \mathbf{P}_1(\mathbf{R}_{22}^T\otimes\mathbf{I}_n)\\
\label{eq:PSI:21}
{\mathbf{\Psi}_{21}} &=&  (\mathbf{R}_{22}\otimes\mathbf{I}_n)\mathbf{P}_2\\
\label{eq:PSI:22}
{\mathbf{\Psi} _{22}} &=& N({\mathbf{R}_{22}}\otimes \mathbf{I}_n)\mathbf{H}({\mathbf{R}_{22}^T}\otimes \mathbf{I}_n) 
\end{eqnarray}
with 
\begin{eqnarray}
\label{eq:Hmatr}
{\mathbf{H}} &:=& (\Vone_{N-1}\Vone_{N-1}^T\otimes\mathbf{A}_1)+\bar{\mathbf{A}}\\
\label{eq:P1Matr}
{\mathbf{P}_1} &:=&  [{\mathbf{A}_2} - {\mathbf{A}_1}, \cdots, {\mathbf{A}_{N}} - {\mathbf{A}_1}]\\
\label{eq:P2Matr}
{\mathbf{P}_2} &:=&   [{\mathbf{A}_2^T} - {\mathbf{A}_1^T}, \cdots, {\mathbf{A}_{N}^T} - {\mathbf{A}_1^T}]^T
\end{eqnarray}

\item the matrix $\mathbf{T}_I=N\mathbf{R}_{22}(\Vone_{N-1}\Vone_{N-1}^T+\mathbf{I}_{N-1})\mathbf{Q}_{22}^T$ was obtained using Lemma \ref{prop:dif_lap} for $(\mathbf{R}^{-1}\otimes \mathbf{I}_n)\widehat{\Li}_I(\mathbf{R}\otimes \mathbf{I}_n)$. 

\item $\widehat{\Lh}:=({\mathbf{R}^{ - 1}}\otimes\mathbf{I}_n) \Lh ({\mathbf{R}}\otimes\mathbf{I}_n)$ and using again Lemma \ref{prop:dif_lap} yields 
$$
\widehat{\Lh} = \left[ {\begin{array}{*{20}{c}}
{0}&{\Vzero_{1\times (N-1)}}\\
{\Vzero_{(N-1)\times 1}}&{\sigma \bar{\mathbf{\Lambda}}_C + \sigma_P \mathbf{T}_P\bar{\mathbf{\Lambda}}_P\mathbf{T}_P^T}
\end{array}} \right]\otimes \mathbf{I}_{n}
$$
with $\mathbf{T}_P=N\mathbf{R}_{22}(\Vone_{N-1}\Vone_{N-1}^T+\mathbf{I}_{N-1})\mathbf{U}_{22}^T$.
\end{itemize}
%
%
\subsection{Main Result}
\begin{thm}
\label{Th:I:PI}
Consider the multiplex network \eqref{eq:DMPI} associated to the multigraph $\mathscr{M}= \{\mathscr{G}_{C},\mathscr{G}_{P},\mathscr{G}_{I}  \}$. Assuming the open-loop network structure $\mathscr{G}_{C}$ is connected,  admissible consensus is achieved if the following conditions hold
\begin{enumerate}
{
	\item[i)] The matrix ${\mathbf{\Psi} _{11}} = (1/N)\sum\nolimits_{k = 1}^N {{\mathbf{A}_k}}$ is non-singular, and its symmetric part ${\mathbf{\Psi} _{11}^{\prime}}$ is Hurwitz,
	
	\item[ii)] $\sigma_P\lambda_2({\Lp_P}) >\frac{1}{2}\left( {\frac{\mu }{{N\left| \eta  \right|}} + \rho } \right) -\sigma\lambda_2({\Lc_C})$
	
	\item[iii)] $\lambda_2({\Lp_I}) > 0$ and $\sigma_I>0$ 
}	
\end{enumerate}
where
\begin{subequations}
\label{eq:cond}
\begin{alignat}{3}
\label{eq:conda}
\mu &:= \lambda_{\max}\left( { {{{\sum\nolimits_{k = 2}^N {\left( {{\mathbf{A}_k^{\prime}} -  {{\mathbf{A}_1^{\prime}} } } \right)} }^2}} } \right)\\
\label{eq:condb}
\eta &:= \lambda_{\max}\left(\mathbf{\Psi} _{11}^{\prime}\right)\\
\label{eq:condc}
\rho &:= \mathop {\max }\limits_{k \in \mathcal{N}} \left\{ {{\lambda _{\max }}\left( {{\mathbf{A}_k^{\prime}}} \right)} \right\}
\end{alignat}
\end{subequations}
Moreover, all node states asymptotically converge to $\mathbf{x}_\infty= - (1/N){\mathbf{\Psi}_{11}^{ - 1}}\sum\nolimits_{k = 1}^N {{\mathbf{b}_k}}$.
\end{thm}
\begin{pf*}{Proof.} From the assumptions, ${\mathbf{\Psi} _{11}}$ is a non-singular matrix; therefore, we have that the consensus equilibrium \eqref{eq:equili} exists. Then, consider the candidate Lyapunov function (in what follows we remove the time dependence of the state variables to simplify the notation)
\begin{equation}
\label{lyap:PI}
V = \frac{1}{2}( {{\mathbf{e}}}_1^T{{ {{\mathbf{e}}}}_1} + {{ {\bar{\mathbf{e}}}}^T} {\bar{\mathbf{e}}}) + \frac{1}{{2\sigma_I }}{{ {\bar{\mathbf{w}}}}^T}{(\mathbf{T}_I {{\bar {\mathbf{\Lambda}} }_I}{\mathbf{T}_I^T}\otimes \mathbf{I}_n)}^{-1} {\bar{\mathbf{w}}}
\end{equation}
From Lemma \ref{prop:dif_lap} we know that $\mathbf{T}_I\bar{\mathbf{\Lambda}}_I\mathbf{T}_I^T$ is an eigendecomposition of a symmetric matrix with positive eigenvalues, which  are the diagonal entries of $\bar {\mathbf{\Lambda}}_I$; therefore, its inverse exist and it is also a positive definite matrix. Consequently, \eqref{lyap:PI} is a positive definite and radially unbounded function. Then, differentiating $V$ along the trajectories of \eqref{eq:final:PID} and using expressions \eqref{eq:PSI:12} and \eqref{eq:PSI:21}, one has 
\begin{equation}
\begin{split}
\label{diff:lyap:PI}
\dot V &=  V_1({\mathbf{e}}_1) + V_2(\bar{\mathbf{e}}) + V_3(\bar{\mathbf{e}}) + V_4({\mathbf{e}}_1,\bar{\mathbf{e}})
\end{split}
\end{equation}
where, $V_1({\mathbf{e}}_1) = {\mathbf{e}}_1^T{\mathbf{\Psi} _{11}}{{ {\mathbf{e}}}_1}$, $V_2(\bar{\mathbf{e}}) = \bar {\mathbf{e}}^T \mathbf{\Psi} _{22} \bar{\mathbf{e}}$, $ V_3(\bar{\mathbf{e}}) = - \bar {\mathbf{e}}^T( \sigma (\bar {\mathbf{\Lambda}}_C \otimes \mathbf{I}_n) + \sigma_P (\mathbf{T}_P\bar {\mathbf{\Lambda}}_P\mathbf{T}_P^T\otimes \mathbf{I}_n)) \bar{\mathbf{e}} $, and $V_4({\mathbf{e}}_1,\bar{\mathbf{e}}) = { {\mathbf{e}}_1^T(\mathbf{P}_1+\mathbf{P}_2^T)(\mathbf{R}_{22}^T\otimes \mathbf{I}_n)\bar {\mathbf{e}} }$. Now, we proceed to find an upper-bound for each of the terms in \eqref{diff:lyap:PI}. 
From the assumptions we know that $\mathbf{\Psi} _{11}+\mathbf{\Psi} _{11}^T$ is Hurwitz; therefore, using \eqref{eq:condb} and property \eqref{symm:bound}, one has that $V_1({\mathbf{e}}_1)\leq -(1/2)\left| \eta  \right|{\mathbf{e}}_1^T{\mathbf{e}}_1$. 

{Next, consider the symmetric matrix $\mathbf{\Psi}^{\prime}:=\mathbf{\Psi}+\mathbf{\Psi}^T$; therefore, using \eqref{transpose_R} $\mathbf{\Psi}^{\prime}=({\mathbf{R}^{ - 1}}\otimes\mathbf{I}_n)(\widehat{\mathbf{A}}+\widehat{\mathbf{A}}^T)({\mathbf{R}}\otimes\mathbf{I}_n)$. Then, it immediately follows that ${\lambda _{\max }}\left( {\mathbf{\Psi}+\mathbf{\Psi}^T} \right)=\rho$, where $\rho$ is given in \eqref{eq:condc}. Now, we can write $V_2(\bar{\mathbf{e}}) = (1/2)\bar {\mathbf{e}}^T \mathbf{\Psi} _{22}^{\prime} \bar{\mathbf{e}}$, and from the fact that $\mathbf{\Psi} _{22}^{\prime}$ is a principal sub-matrix of $\mathbf{\Psi}^{\prime}$, by using property \eqref{eig_bloksim} one has $V_2(\bar{\mathbf{e}})\le \rho/2 {{\bar {\mathbf{e}}}^T}{{\bar {\mathbf{e}}}}$.}

From Lemma \ref{prop:dif_lap} we know that $\mathbf{T}_P\bar {\mathbf{\Lambda}}_P\mathbf{T}_P^T$ is a symmetric positive definite matrix. Hence, using \eqref{symm:bound} we have that $V_3(\bar{\mathbf{e}}) \leq -(\sigma \lambda_2({\Lc_C})  + \sigma_P\lambda_2({\Lp_P}) ) {{ \bar{\mathbf{e}}}^T} {{ \bar{\mathbf{e}}}}$. 

Finally, setting $\mathbf{v}_1=\mathbf{e}_1$, $\mathbf{v}_2=\bar{\mathbf{e}}$, $\mathbf{Q}_1^T=\mathbf{P}_1+\mathbf{P}_2^T$ and $\mathbf{Q}_2=\mathbf{R}_{22}^T\otimes \mathbf{I}_n$ and using \eqref{pos:eq:1} yields
\[
\begin{split}
V_4({\mathbf{e}}_1,\bar{\mathbf{e}}) &< \frac{\varepsilon}{2}{\mathbf{e}}_1^T\mathbf{Q}_1^T\mathbf{Q}_1{\mathbf{e}}_1 + \frac{1}{2\varepsilon}\bar {\mathbf{e}}^T\mathbf{Q}_2^T\mathbf{Q}_2\bar {\mathbf{e}}\\
& < {\frac{\varepsilon}{2}{\mathbf{e}}_1^T{ {{{\sum\limits_{k = 2}^N {\left( {{\mathbf{A}_k^{\prime}} -  {{\mathbf{A}_1^{\prime}} } } \right)} }^2}} }{\mathbf{e}}_1+ \frac{1}{2\varepsilon}\bar {\mathbf{e}}^T\mathbf{Q}_2^T\mathbf{Q}_2\bar {\mathbf{e}} }
\end{split}
\]
We can further simplify this expression by noticing that $\mathbf{Q}_2^T\mathbf{Q}_2$ is a symmetric matrix and using \eqref{symm:bound}, \eqref{specnorm:bou}, and \eqref{prop:U:5norm}, we can write $\bar{\mathbf{e}}^T\mathbf{Q}_2^T\mathbf{Q}_2\bar {\mathbf{e}} \le \specnorm{\mathbf{Q}_2}^2\bar{\mathbf{e}}^T\bar {\mathbf{e}}  \le (1/N)\bar{\mathbf{e}}^T\bar {\mathbf{e}}$. Then, using \eqref{eq:conda} yields ${V_4({\mathbf{e}}_1,\bar{\mathbf{e}})} \le ({\varepsilon\mu})/{2}{\mathbf{e}}_1^T{\mathbf{e}}_1 + {1}/{(2N\varepsilon)}\bar {\mathbf{e}}^T\bar {\mathbf{e}}$. Exploiting all the bounds we found for each term in \eqref{diff:lyap:PI} yields
\begin{equation}
\label{diff:lyap:PIb}
\begin{split}
\dot V &\le (1/2)\left(\varepsilon\mu-\left| \eta  \right|\right){\mathbf{e}}_1^T{{ {\mathbf{e}}}_1} - (\sigma \lambda_2(\Lc_C)+\sigma_P \lambda_2(\Lp_P) ){{\bar {\mathbf{e}}}^T}{{\bar {\mathbf{e}}}} \\
& \quad  + \left( {\frac{1}{2N\varepsilon} +  \frac{\rho}{2}} \right) \bar {\mathbf{e}}^T\bar {\mathbf{e}}\\
&\le \xi_1{\mathbf{e}}_1^T{{ {\mathbf{e}}}_1} + \xi_2 \bar {\mathbf{e}}^T\bar {\mathbf{e}}
\end{split}
\end{equation}
where $ \xi_1:=\varepsilon\mu-\left| \eta  \right|<0$ and $\xi_2:=1/(2N\varepsilon)+\rho/2 - \sigma \lambda_2(\Lc_C)- \sigma_P \lambda_2(\Lp_P) <0$. Now, $\xi_1<0$ is ensured if $\varepsilon<\left| \eta  \right|/\mu$. Also, $\xi_2<0$ if condition ii) is fulfilled. Therefore, under the hypotheses, all agents in \eqref{eq:sys:1} achieve admissible consensus to $\mathbf{x}_{\infty}$ as defined in \eqref{eq:equili}.
\end{pf*}
\begin{rem}\hspace{0.2cm}
\begin{itemize}
\item Note that the conditions of Theorem 11 can be used as an effective tool to tune the control gain and/or rewire the control layers.



%
\item The stability analysis problem for the whole network has been simplified. In particular, rather than studying the stability of the  $2nN\times 2nN$ matrix in \eqref{eq:DMPI}, only conditions  i) and ii) need to be verified which only depend upon $n\times n$ matrices.
\item Note that condition (ii) can always be ensured by choosing $\sigma_P$ sufficiently large. Crucially, our bound, depending on the network structure and the node dynamics, allows to estimate the threshold value of $\sigma_P$ required to guarantee global convergence. This can be extremely useful when tuning the gains in practice and also for network design.
\item It is important to highlight that optimal values for the proportional layer ($\sigma_P$, $\lambda_2({\Lp_P})$) can be obtained by properly labeling node 1 so that $\mu$ is such that the quantity ${\mu }/({{N\left| \eta  \right|}})$ in condition ii) is the smallest. 
\item The topology of the integral control layer can be chosen arbitrarily. Hence, the independence of its structure from that of the other layers allows to minimize the number of control interventions across the network. 
%
\end{itemize}
\end{rem}
%
%
In the case where the graph associated to the open loop network $\mathcal{L}_C$ is connected, it is possible to use the following result that comes immediately from Theorem 11.
\begin{cor} Let $\mathscr{G}_{cp}=\mbox{proj}(\mathscr{G}_C,\mathscr{G}_P)$ denotes the projection graph of $\mathscr{G}_{C}$ and $\mathscr{G}_{P}$ and $\boldsymbol{\mathcal{L}}_{cp}$ be its associated Laplacian matrix; then, assuming $\mathscr{G}_{cp}$ is connected, the multiplex closed-loop network \eqref{eq:DMPI}  reaches admissible consensus if conditions i) and iii) of Theorem \ref{Th:I:PI} are fulfilled together while condition ii) is substituted with $\lambda_2(\boldsymbol{\mathcal{L}}_{cp}) >({1}/{2})\left( {{\mu }/{({N\left| \eta  \right|})} + \rho } \right)$. 
\end{cor}


\begin{pf*}{Proof.} Since the graph $\mathscr{G}_{cp}=\mbox{proj}(\mathscr{G}_{C},\mathscr{G}_{P})$ is connected then we have that $\boldsymbol{\mathcal{L}}_{cp}=\mathbf{U}\mathbf{\Lambda}_{cp}\mathbf{U}^T$ where $\mathbf{U}$ is the matrix composed by the eigenvectors of $\boldsymbol{\mathcal{L}}_{cp}$ and $\mathbf{\Lambda}_{cp}=\mbox{diag}\{0,\lambda_2(\boldsymbol{\mathcal{L}}_{cp}),\cdots,\lambda_N(\boldsymbol{\mathcal{L}}_{cp}) \}$. Hence, we have that $\Lh=({\boldsymbol{\mathcal{L}}_{cp} \otimes {\mathbf{I}_n}})$ in \eqref{eq:DMPI} and following a similar procedure as in Section \ref{Sec:conv} completes the proof.
\end{pf*}
%
%
\begin{cor} Considering a connected open-loop network with homogeneous node dynamics, i.e $\mathbf{A}_i=\mathbf{A}, i\in \mathcal{N}$ where $\mathbf{A}$ and $\mathbf{A}^{\prime}$ are Hurwitz stable. Then the closed-loop network \eqref{eq:DMPI}, reaches admissible consensus 
for any connected proportional and integral graph topologies with $\sigma_P,\sigma_I>0$.
\end{cor}
\begin{pf*}{Proof.}
Firstly, note that when all nodes share the same intrinsic dynamics we have that $\mu=0$ in \eqref{eq:conda}, and ${\mathbf{\Psi} _{11}}=\mathbf{A}$. Hence, from the assumptions, conditions i) and iii) of Theorem \ref{Th:I:PI} are automatically satisfied and from the fact that matrix $\mathbf{A}+\mathbf{A}^T$ is Hurwitz, one has that $\rho<0$ in \eqref{eq:condc}; therefore, condition ii) of Theorem \ref{Th:I:PI} is also automatically fulfilled.
\end{pf*}
Now consider the case where ${\mathbf{\Psi} _{11}}$ is not Hurwitz stable; then, it is possible to apply a local feedback control action to a subset of the nodes so as to render ${\mathbf{\Psi} _{11}}$ Hurwitz stable and guarantee the existence of the consensus equilibrium $(\mathbf{x}^*,\mathbf{z}^*)$ in the closed-loop network. Or, equivalently, make the network \textit{consensuable} according to the definition given in \cite{Wang2014}. Specifically, consensusability can be achieved by adding an extra control input, say $\mathbf{v}_i$, onto a fraction $K<N$ nodes so that $\mathbf{\Psi}_{11}$ is stable. For example, one can choose the controller
\begin{equation}
\label{control:II}
{\mathbf{v}_i}(t) = \mathbf{H}_i{\mathbf{x}_i}(t) 
\end{equation}
where $\mathbf{H}_i\in \amsbb{R}^{n\times n}$ is a gain matrix to be designed appropriately. Note that typically one could simply choose $K=1$ so that the dynamics of just one node is altered by this feedback controller.
\begin{cor} The heterogeneous network \eqref{eq:sys:1} is said to be consensusable under the distributed control action \eqref{control:II}, if there exist matrices $\mathbf{H}_i$ such that conditions i), ii) and iii) in Theorem \ref{Th:I:PI} are fulfilled.
\end{cor}
\begin{rem} Note that the presence of local controllers acting on some nodes can be used not only for improving the closed-loop network stability, but also to change the value of the consensus vector $\mathbf{x}_{\infty}$.   
\end{rem}
\subsection{Control Algorithm}
\label{cont:alg}
The results presented so far can be distilled into the following algorithmic steps to design the multilayer PI network control strategy proposed in this paper. Specifically,
\begin{enumerate}
\item[S1] 
Compute matrix ${\mathbf{\Psi} _{11}}=(1/N)\sum\nolimits_{k = 1}^N {{\mathbf{A}_i}}$ from the open-loop network \eqref{eq:sys:1}. 

\item[S2] 
If matrix  ${\mathbf{\Psi} _{11}}$ and ${\mathbf{\Psi} _{11}^{\prime}}$ are Hurwitz stable then go to step S4, otherwise go to S3.

\item[S3] 
Design local controllers \eqref{control:II} such that ${\mathbf{\Psi} _{11}}$ together with its symmetric part ${\mathbf{\Psi} _{11}^{\prime}}$ are Hurwitz. Note that matrices $\mathbf{H}_i$ can also be properly chosen for selecting different values of the consensus vector $\mathbf{x}_\infty$ in \eqref{eq:equili}

\item[S4] Select any connected and weighed undirected graph $\mathscr{G}_{I}$ for the integral layer e.g. a minimal spanning tree. Then compute the quantities $\mu$, $\eta$, and $\rho$ defined in \eqref{eq:cond}

\item[S5] Find a connected and weighed undirected graph $\mathscr{G}_{P}$ for the proportional layer and a value of the global coupling gain $\sigma_P$ such that $\sigma_P\lambda_2({\Lp_P}) >(1/2)\left( {{\mu }/ ({{N\left| \eta  \right|}}) + \rho } \right) - c\lambda_2({\Lc_C})$

\end{enumerate}

\subsection{Example}
\label{ExampleI}
For the sake of simplicity and without loss of generality we consider three types of node dynamics; oscillatory ($\mathbf{E}_1$), stable ($\mathbf{E}_2$) and unstable ($\mathbf{E}_3$)
\[{\mathbf{E}_1}: = \left[ {\begin{array}{*{20}{c}}
{0}&1\\
-1&{0}
\end{array}} \right],{\mathbf{E}_2}: = \left[ {\begin{array}{*{20}{c}}
-1.5&0\\
{-1}&-1
\end{array}} \right],{\mathbf{E}_3}: = \left[ {\begin{array}{*{20}{c}}
1&1\\
0&{0.5}
\end{array}} \right]\]
Then, we consider eight decoupled agents governed by \eqref{eq:sys:1}, with $\sigma = 0$, $\mathbf{A}_k = \mathbf{E}_1, k \in \{1,3\}$, $\mathbf{A}_k = \mathbf{E}_2,k \in \{2,5,7\}$, and $\mathbf{A}_k = \mathbf{E}_3, k \in \{4,6,8\}$ and disturbances $\mathbf{b}_i\in{\amsbb{R}^{2\times 1}}$ given by $\mathbf{B} = \left[\mathbf{b}_1^T,\cdots,\mathbf{b}_8^T\right]^T= \left[0,10,0,30,0,1,20,0,30,30,60,10,-10,40,0,0 \right]$.
Note that no disturbance is acting on the 8-th node and that some of the agents are marginally stable or unstable. Nevertheless, their average dynamics is characterised by a full rank matrix ${\mathbf{\Psi} _{11}}$ so that Proposition 8 ensures the existence of a consensus equilibrium while Theorem 11 can be used to prove convergence under the action of our multiplex PI strategy.

\begin{figure}[tbp]
\centering {
\subfigure[]
{\label{fig:oprna}
{\includegraphics[scale=0.27]{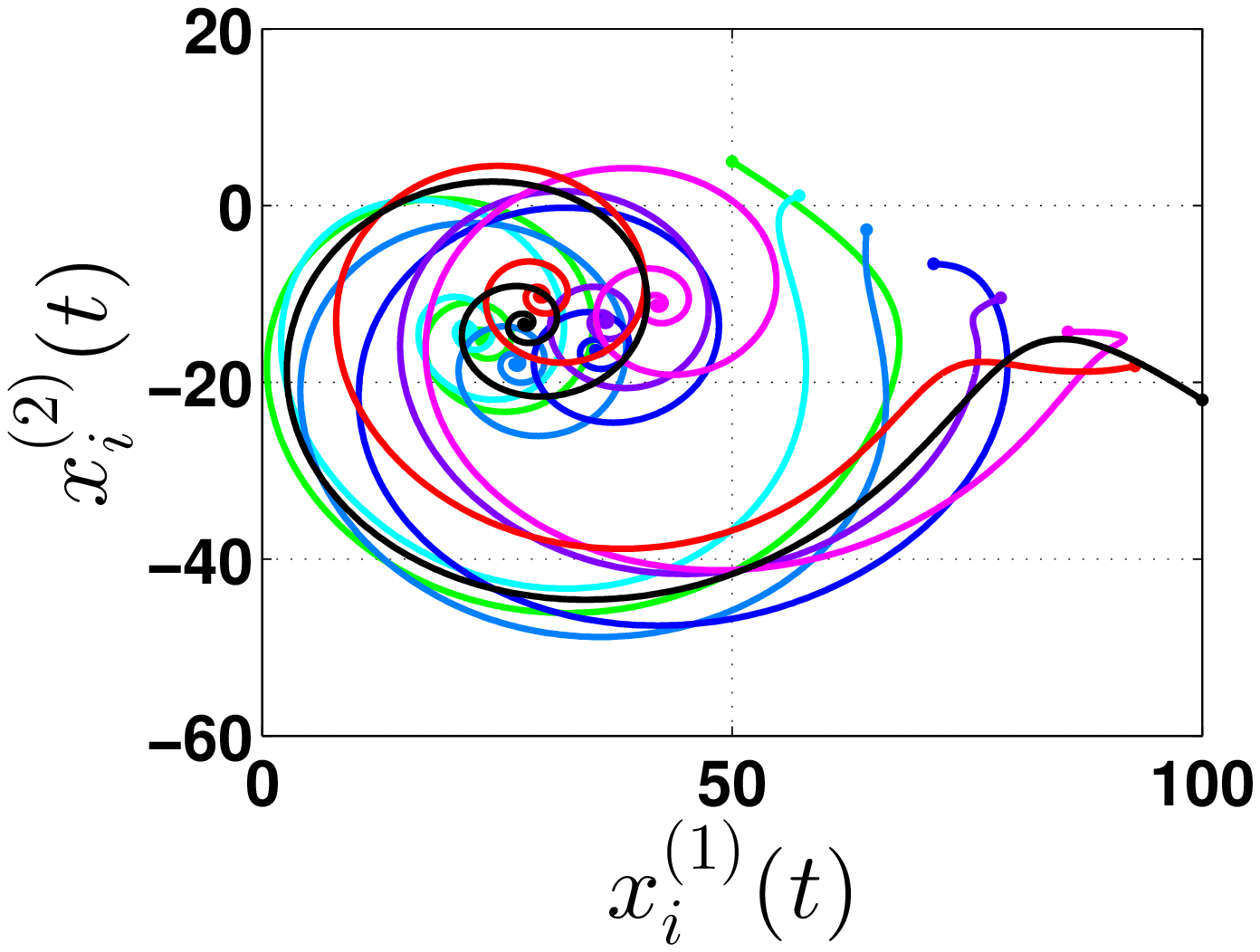}}}
\subfigure[]
{\label{fig:oprnb}
{\includegraphics[scale=0.27]{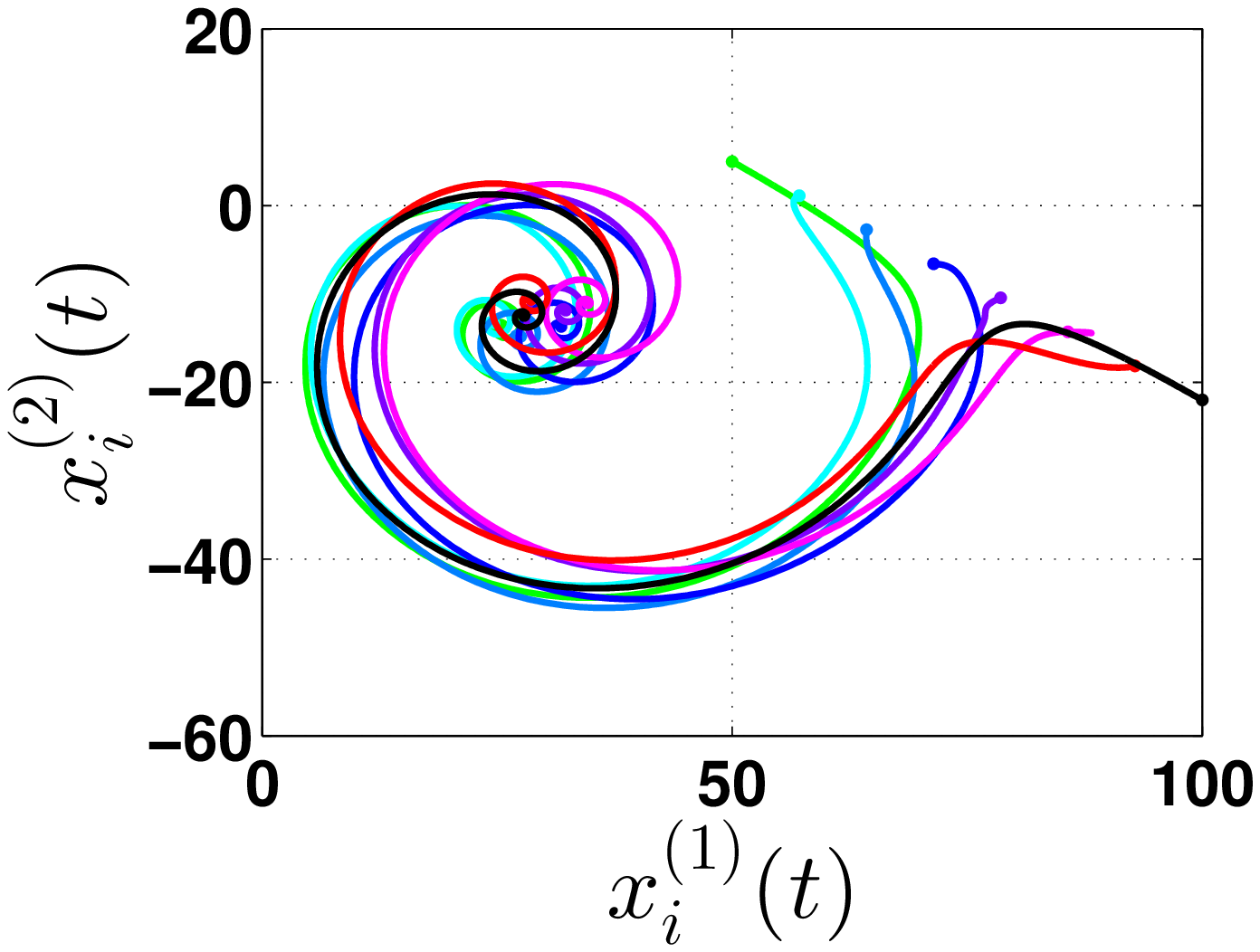}}}
}
  \caption{State space evolution of the heterogeneous network controlled by distributed proportional control for: (a) $\sigma_P=5$ and (b) $\sigma_P=10$.}
  \label{fig:oprn}
\end{figure}

To show the effectiveness of such an approach, for the sake of comparison we start by using a distributed proportional controller setting $\sigma_I=0$ in $\eqref{eq:cont:2b}$. As can be seen in Fig. \ref{fig:oprn}, this can only guarantee bounded convergence. 
To achieve admissible consensus, we deploy next the multiplex PI-Control strategy presented in this paper. Following the control design steps in Section \ref{cont:alg}, we have from S1 that
\[{\mathbf{\Psi} _{11}} = \left[ {\begin{array}{*{20}{c}}
{-0.1875}&{0.625}\\
{-0.625}&{-0.1875}
\end{array}} \right],{\mathbf{\Psi} _{11}^{\prime}} = \left[ {\begin{array}{*{20}{c}}
{ -0.375}&{0}\\
{0}&{-0.375}
\end{array}} \right]\]
where ${\mathbf{\Psi} _{11}}$ is a full rank matrix and ${\mathbf{\Psi} _{11}^{\prime}}$ is a Hurwitz stable matrix. Then, following S4 we select a ring network of 8 nodes with unitary weights ($\beta_{ij}=1 \quad \forall i,j \in \mathcal{N}$) as the connected integral network, and from \eqref{eq:cond} we have that $\mu=59.8328$, $\eta=0.3750$, and $\rho=2.618$. From S5 we have that $\sigma_P\lambda_2({\Lp_P})>11.2812$. Then, choosing, w.l.o.g again a ring network with $\alpha_{ij}=1 \quad \forall i,j \in \mathcal{N}$ so that $\lambda_2({\Lp_P}) = 0.5858$,  the closed-loop network of 8 agents achieves admissible consensus for $\sigma_P>19.25$.

\begin{figure}[tbp]
\centering {
\subfigure[]
{\label{fig:Tim:resa}
{\includegraphics[scale=0.27]{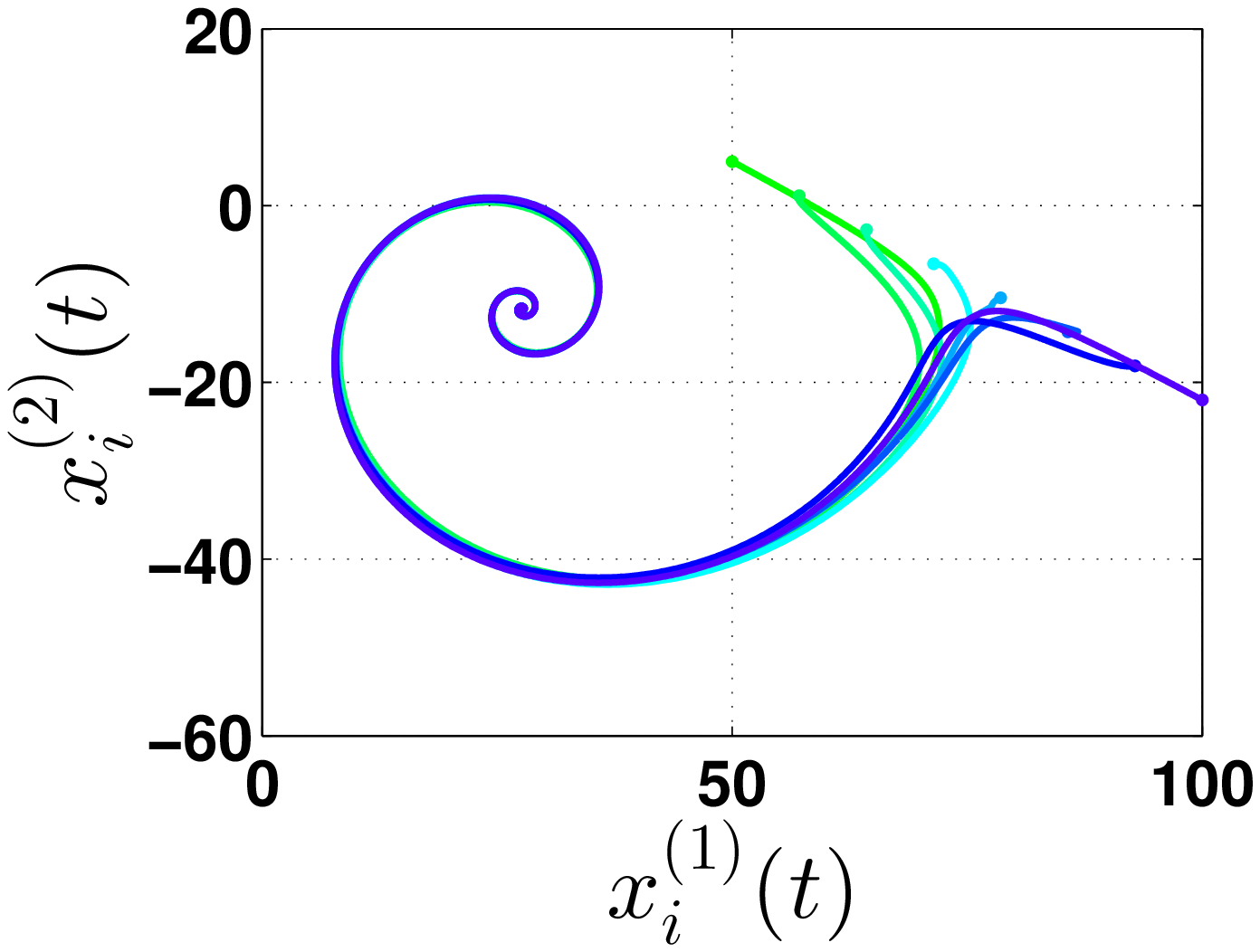}}}
\subfigure[]
{\label{fig:Tim:resb}
{\includegraphics[scale=0.27]{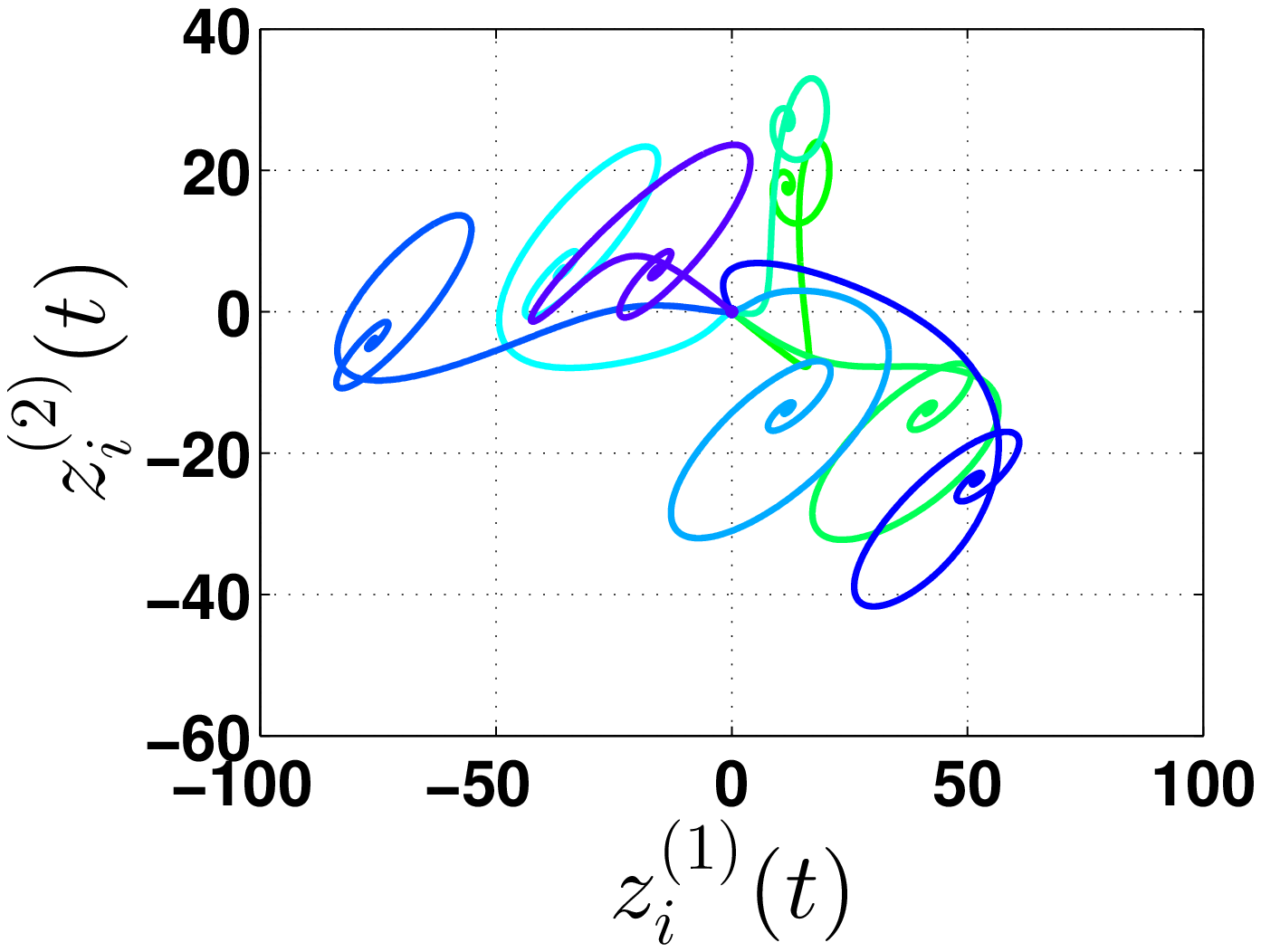}}}
}
  \caption{State space evolution of the closed-loop multiplex network for $\sigma_P=19.3$ and $\sigma_I=15$ where the proportional and integral networks have both ring structures with all weights equal to 3 and 1 respectively.}
  \label{fig:Tim:res}
\end{figure}

We choose $\sigma_P=19.3$, and $\sigma_I=15$. The resulting evolution of the node states and integral actions is shown in Fig. \ref{fig:Tim:res}, where admissible consensus is reached as expected to the predicted value ${\mathbf{x}_\infty }:= - (1/N)\mathbf{\Psi} _{11}^{-1}\sum\nolimits_{k = 1}^N {{\mathbf{b} _k}}= [27.7064, -11.6881]^T$ and the integral terms remain bounded. 
%
%
%
%
\subsection{Discussion}
The admissible consensus conditions presented in Theorem \ref{Th:I:PI} only require the graph structure of the integral layer $\mathscr{G}_{I}$ to be connected. However, in general, we found that the stability of the consensus equilibrium and the rate of convergence are affected by the specific choice of $\mathscr{G}_{I}$.
\begin{figure}[tbp]
\centering {
\subfigure[]
{\label{fig:varNeta}
{\includegraphics[scale=0.15]{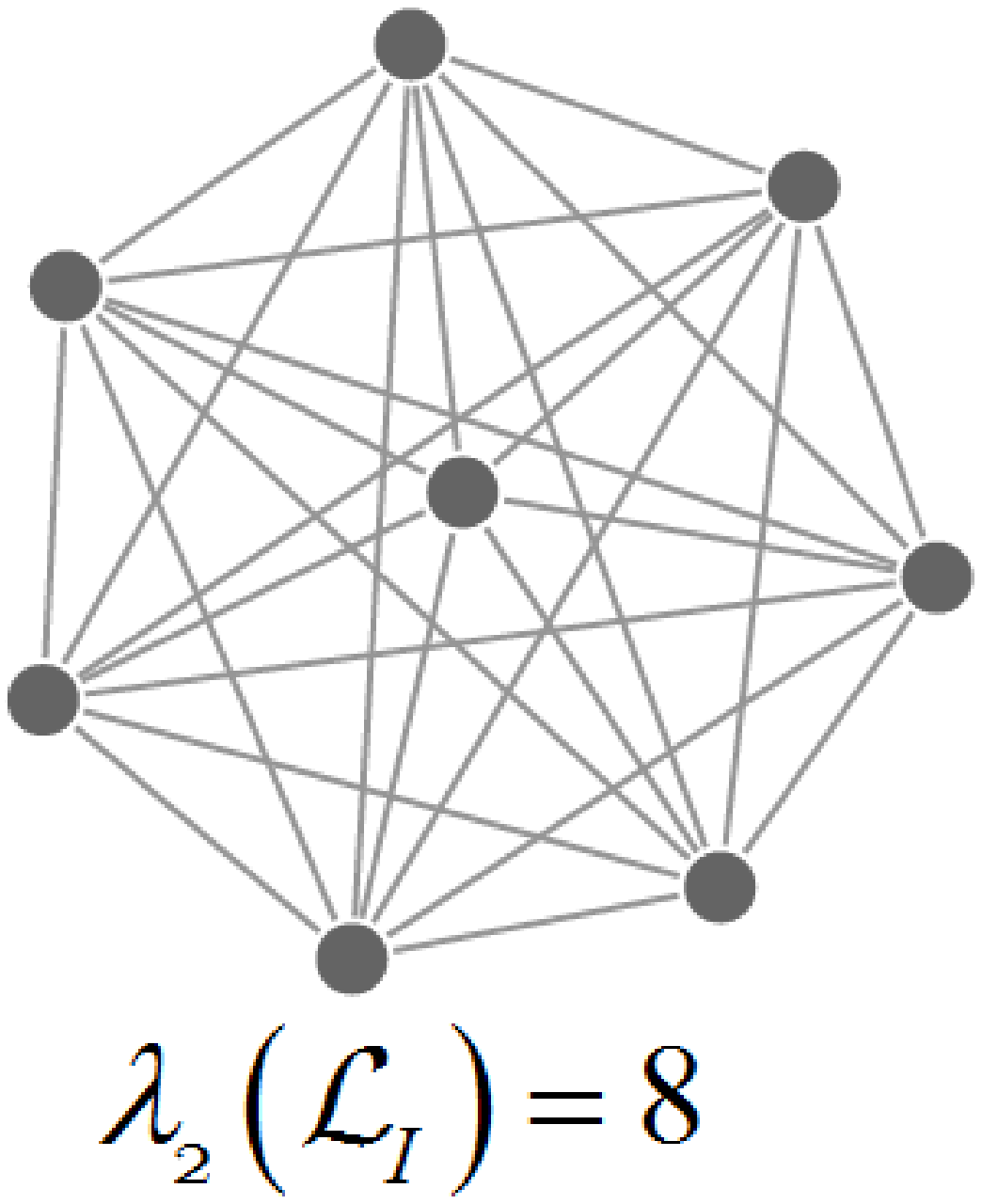}}}
\subfigure[]
{\label{fig:varNetb}
{\includegraphics[scale=0.15]{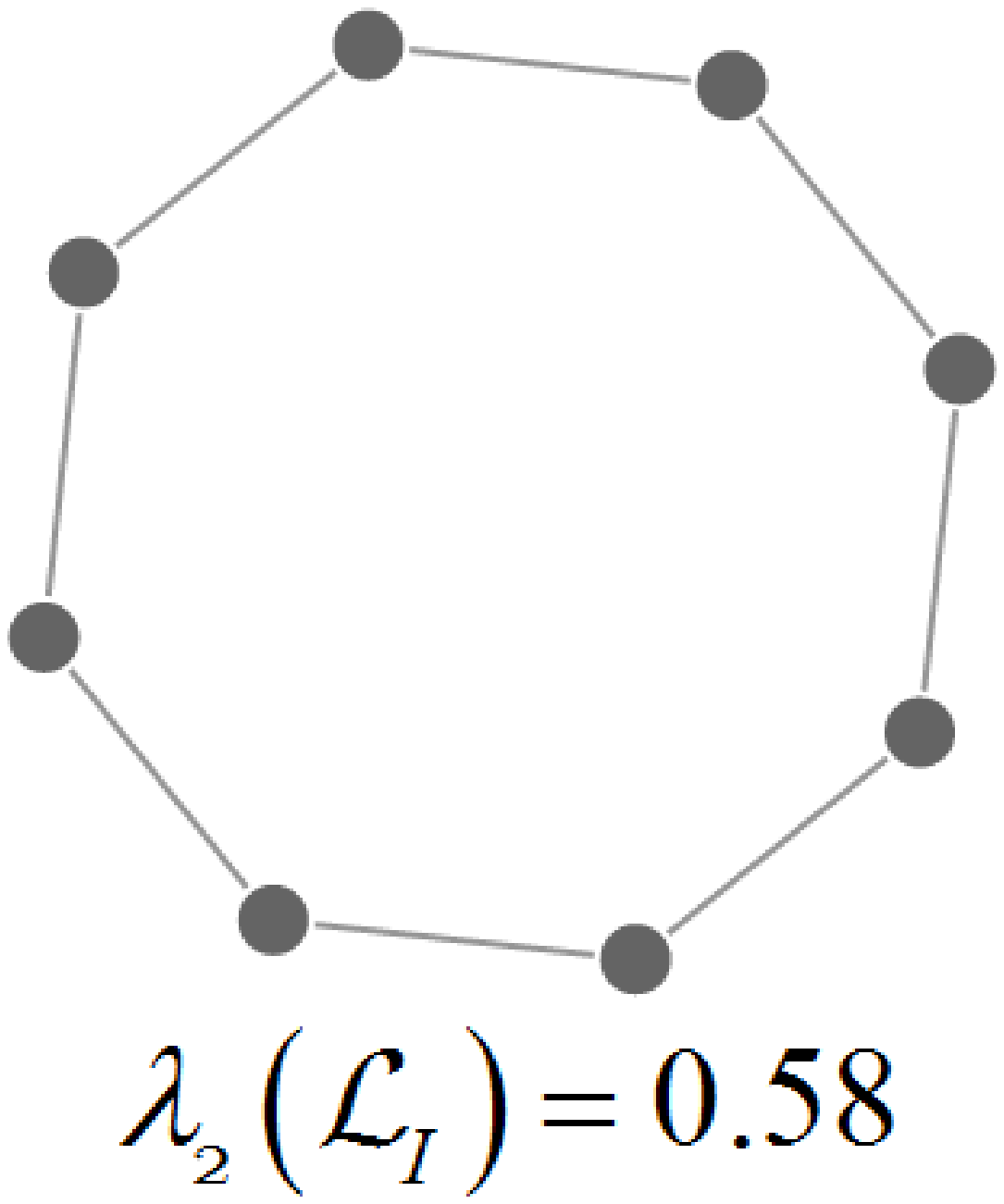}}}
\subfigure[]
{\label{fig:varNetc}
{\includegraphics[scale=0.15]{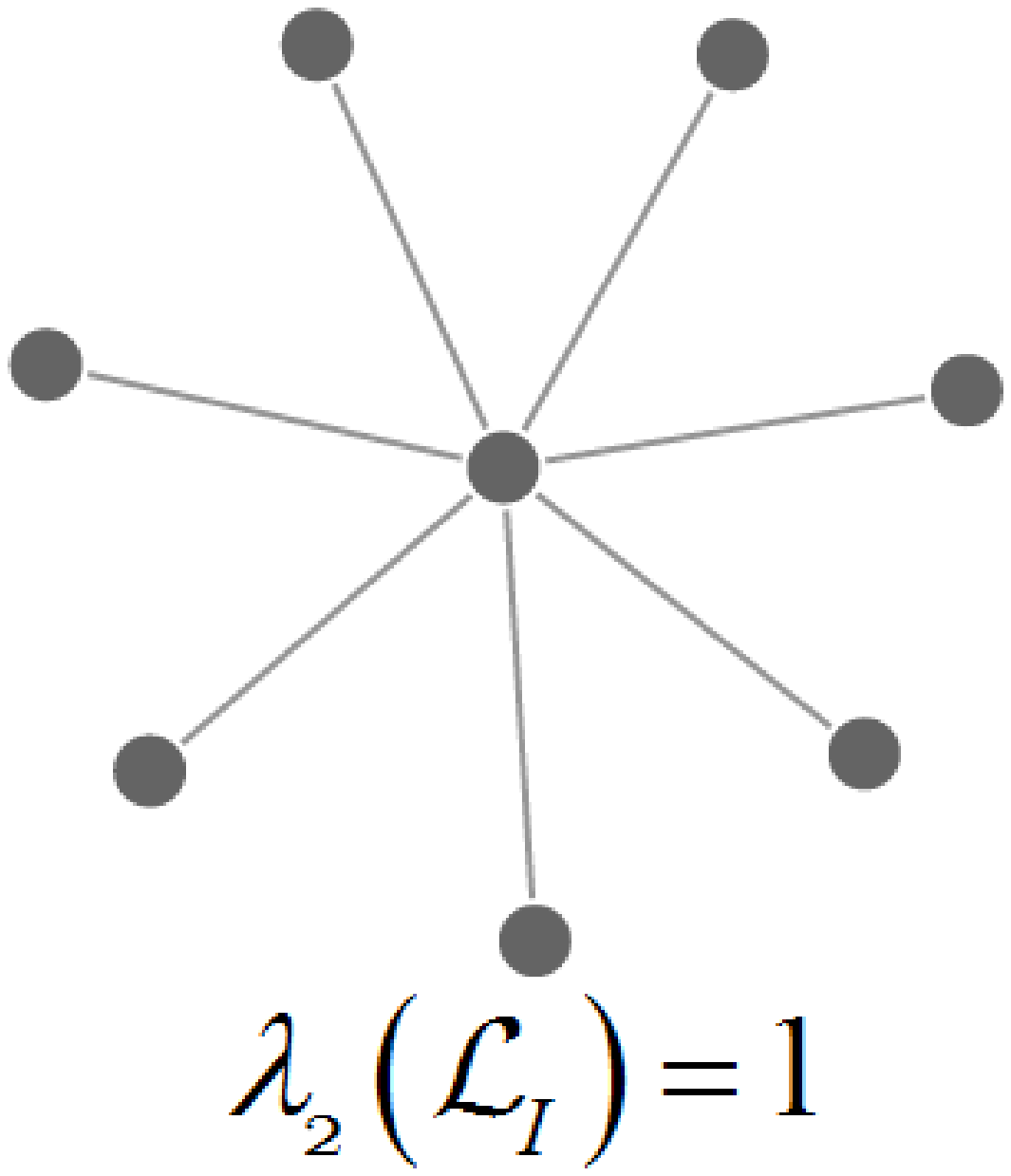}}}
\subfigure[]
{\label{fig:varNetd}
{\includegraphics[scale=0.15]{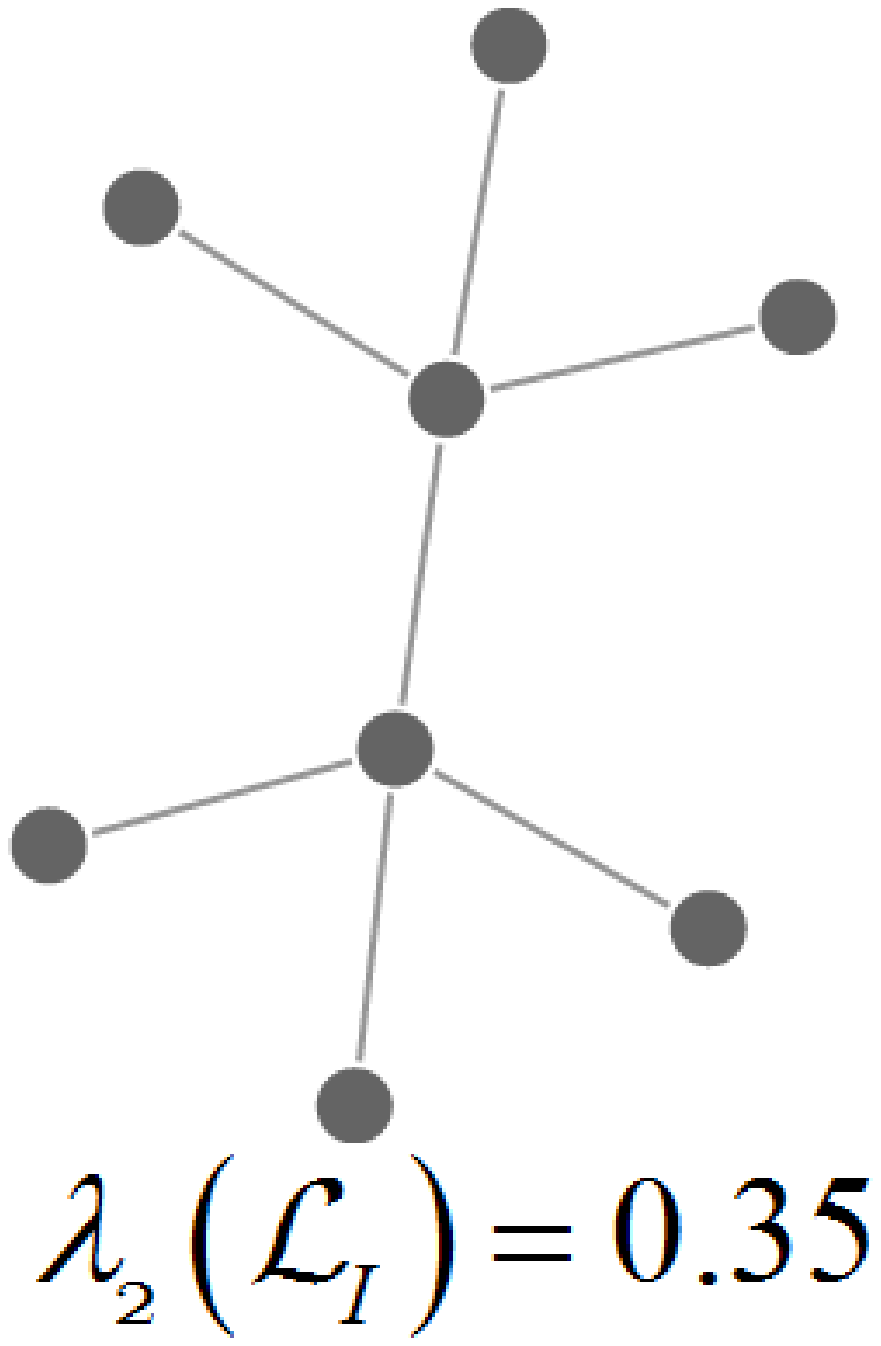}}}
\subfigure[]
{\label{fig:Ring_All2All}
{\includegraphics[scale=0.27]{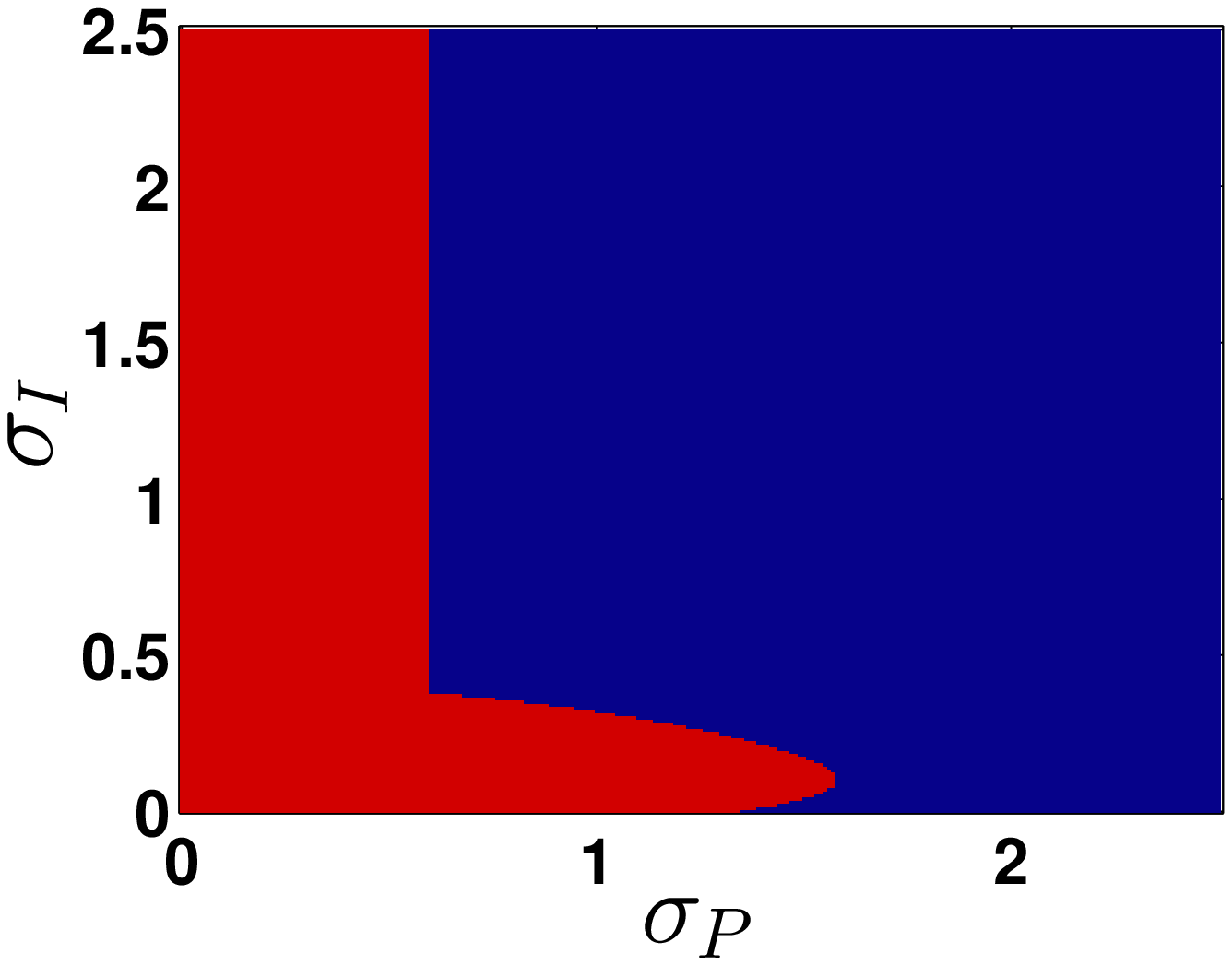}}}
\subfigure[]
{\label{fig:Ring_star}
{\includegraphics[scale=0.27]{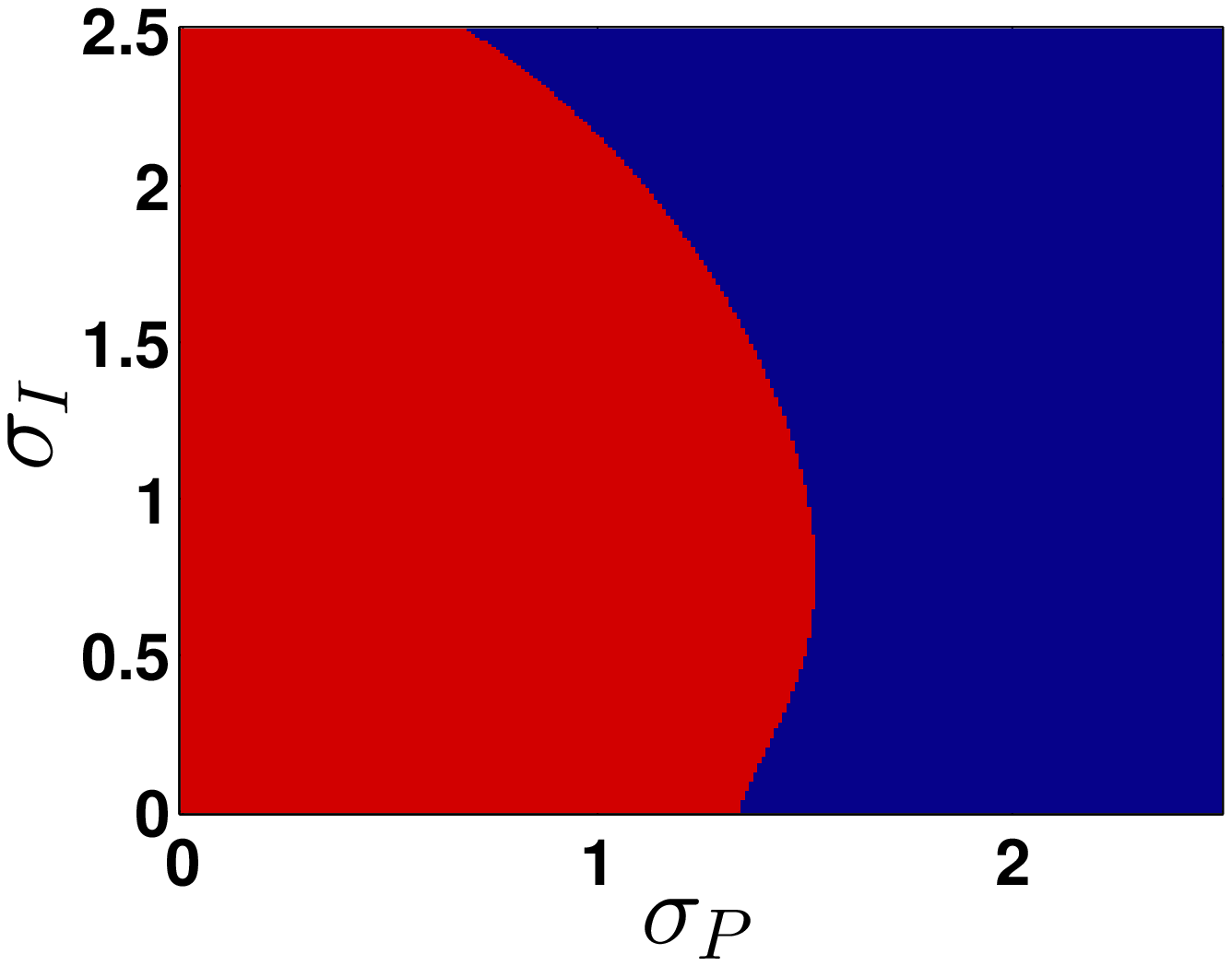}}}
\subfigure[]
{\label{fig:Ring_Ring}
{\includegraphics[scale=0.27]{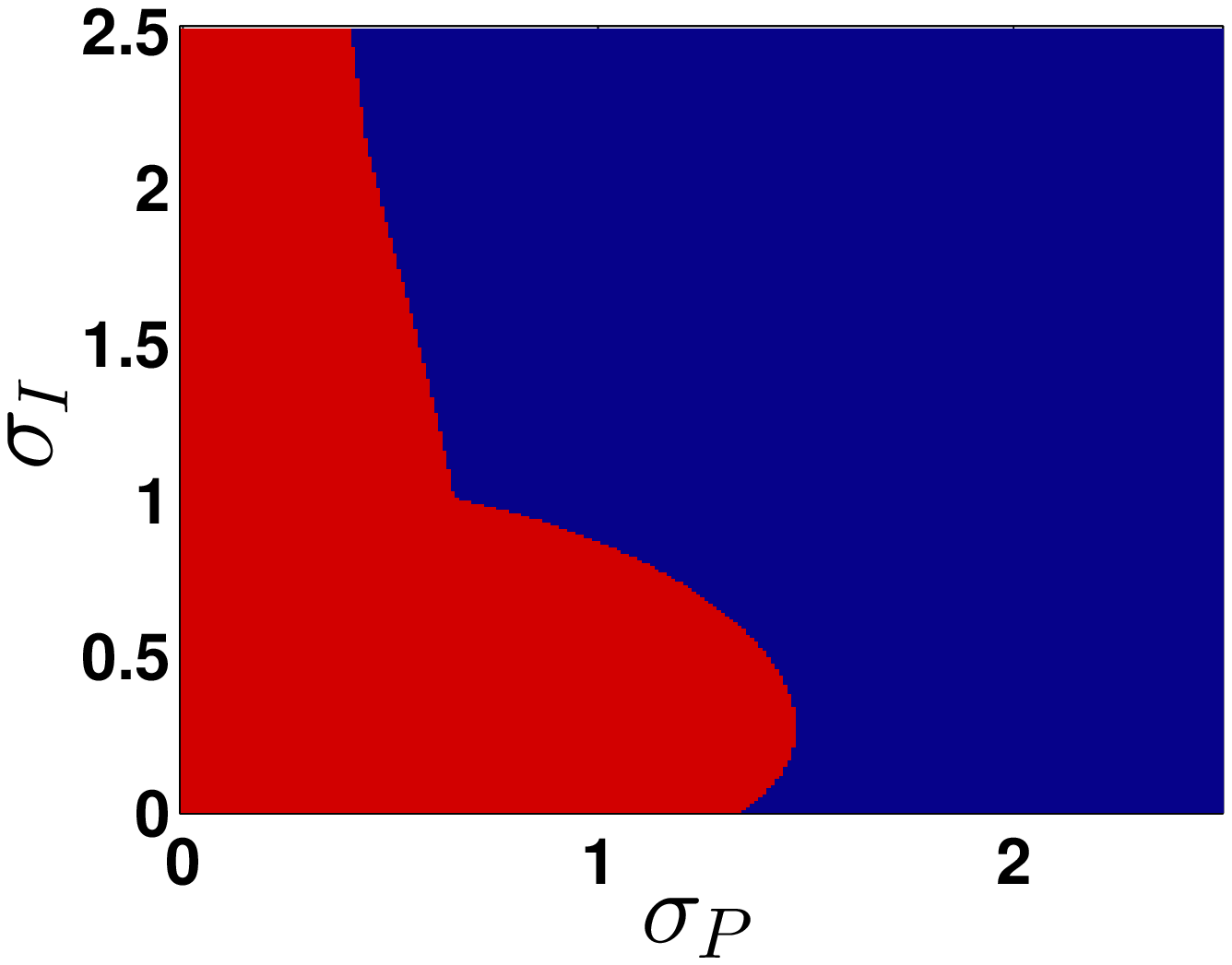}}}
\subfigure[]
{\label{fig:Ring_Tree}
{\includegraphics[scale=0.27]{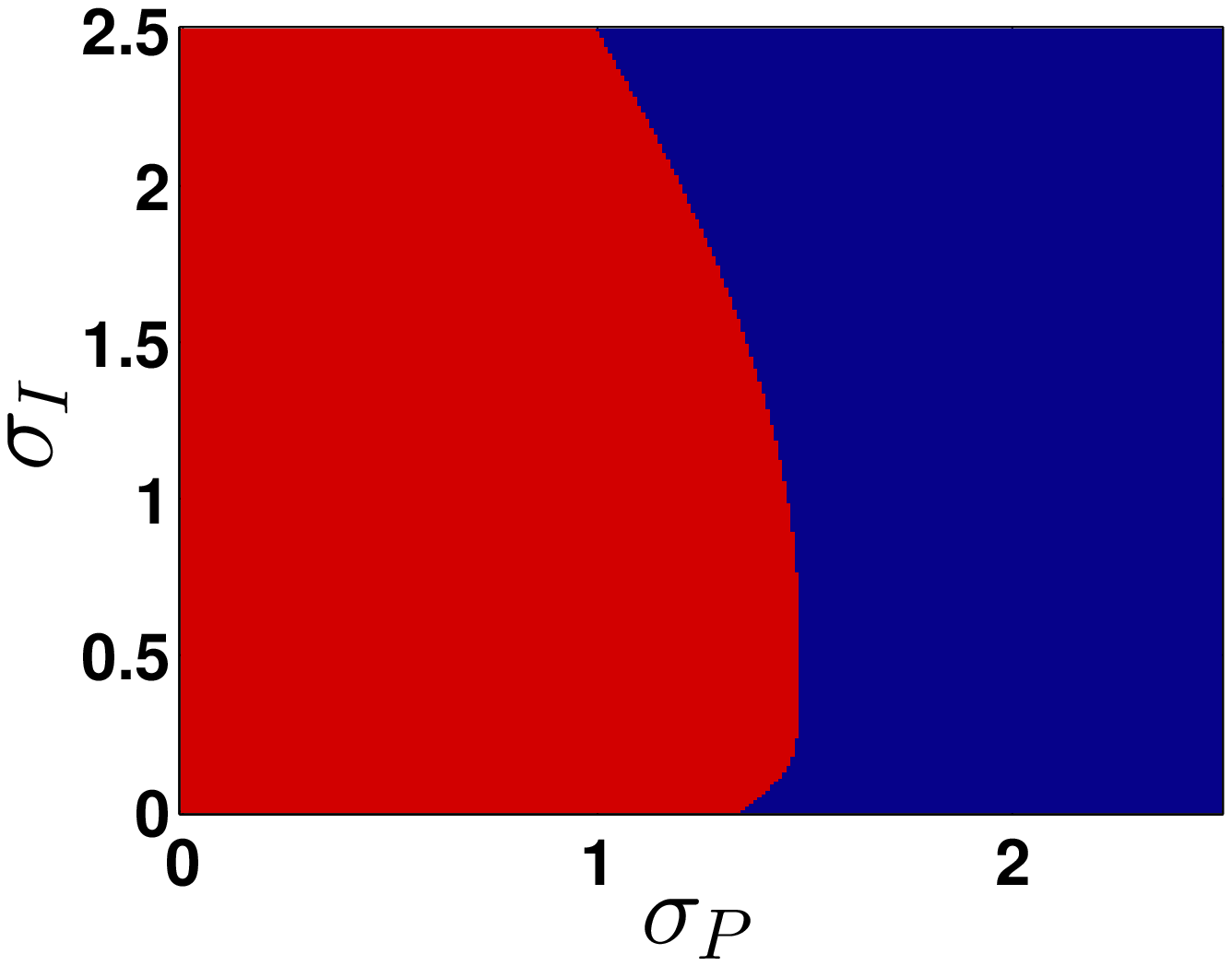}}}
}
\caption{Different network structures with unitary weights considered for the integral control layer: (a) all-to-all, (b) star, (c) ring, and (d) Tree. Two-dimensional stability diagrams varying the topology of $\mathscr{G}_{I}$ [(e): all-to-all, (f) star, (g) ring, (h) tree]. Red regions denote parameter values where consensus is not achieved, blue regions those where consensus is attained.}
\label{fig:StabilRegio} 
\end{figure}
\begin{figure}[tbp]
\centering {
{\includegraphics[scale=0.42]{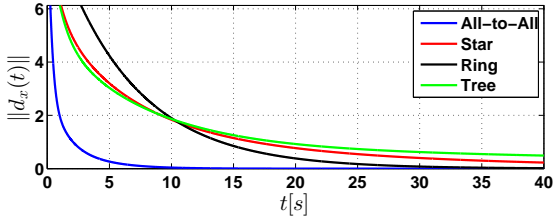}}
}
\caption{Time response of the consensus index $d_x$ when the topology of the integral network is varied}
\label{fig:varNet} 
\end{figure}
To illustrate this point, we considered different structures for the graph $\mathscr{G}_{I}$ while leaving $\mathscr{G}_P$ unchanged, and computed two-dimensional stability diagrams in the control parameter space $(\sigma_P, \sigma_I)$, see Fig. \ref{fig:StabilRegio}.  Namely, at each point in the  $(\sigma_P, \sigma_I)$ space, we computed the maximum eigenvalue of the error system dynamics \eqref{eq:final:PID} depicting in blue those points where the eigenvalue is negative (consensus is achieved) and in red  those where it is positive (convergence is not attained). As shown in Fig. \ref{fig:StabilRegio}(e)-(f), varying the structure of the integral layer has a notable effect on the shape of the stability region. 
%
%
%
We also found that changing the structure of $\mathscr{G}_I$ influences the speed of convergence of the closed-loop multiplex network towards consensus. Specifically, in Fig. \ref{fig:varNet}, we plot the time evolution of the consensus index ${d_x} := \left\| {\mathbf{x}(t) - (1/N)\left( {{\Vone_N}\Vone_N^T \otimes {\mathbf{I}_n}} \right)\mathbf{x}(t)} \right\|$, where $d_x=0$ indicates that the closed-loop network has reached admissible consensus. We observe that the structure of $\mathscr{G}_{I}$ changes the speed of convergence. 
Obtaining an analytical estimate of such a rate is a highly cumbersome task as discussed in \cite{Kenji2011}, but some estimations can be found in the case where the agents are one-dimensional and homogeneous \cite{BurbanoLombana2015}. 
%
%

Finally, it is worth pointing out that in a practical implementation of the multiplex strategy \eqref{eq:cont:2b}, the relative difference $(\mathbf{x}_j(t)-\mathbf{x}_i(t))$ between agents may be affected by measurement errors \cite{Garulli2011,Liu2011,Meng2014}. This might render the integral terms unable to converge. 
In practice, anti-windup strategies (saturations) can be added to the integral terms or higher order actions (e.g. PI$^m$) can be used. 
Also, the multiplex nature of the proposed PI strategy can be further exploited if an estimate of the measurement errors is available. In this case, given that the integral and control layers can have different structures, integral actions can only be deployed on those edges which are less noisy than the others. Preliminary simulations (not reported here for the sake of brevity) confirm this observation which will be the subject of future work.
%
%
\section{Application to Power systems}
In this section, we show that the convergence analysis used to prove stability of the multiplex PI strategy developed in this paper can be effectively used to prove  the emergence of synchronisation in heterogeneous networks of power generators. 
Specifically, we consider $N$ power generators governed by the swing equation \cite{Motter2013power}
\begin{equation}
\label{swing:eq}
\frac{{2{H_i}}}{{{\omega _R}}}{\ddot \delta _i} = P_i^m(t) - P_i^{net}(t), i\in \mathcal{N} 
\end{equation}
where $H_i$ and $\omega _R$ are constants representing the inertia and reference frequency for the $i$-th generator. The quantity $P_i^m(t):=P_i^*-d_i{\dot \delta _i}(t)$ is the mechanical power provided by the $i$-th generator and it is composed by a constant power injection $P_i^*$ and a damping term $d_i{\dot \delta _i}(t), d_i>0$ which models power losses and primary control loops. Moreover, $P_i^{net}(t)$ is the power demanded by the network. Note that when \eqref{swing:eq} is at rest onto an equilibrium, $P_i^m = P_i^{net}$ and the frequency of each generator ${\omega _i}(t): = {\dot \delta _i}(t)$ remains equal to a common constant for all generators in the grid.
For the sake of simplicity, we linearize the swing equation \eqref{swing:eq} around the synchronous state ${\omega _1}(t) =  \cdots  = {\omega _N}(t)$, letting $m_i= 2H_i/\omega _R$, we obtain \cite{Andreasson2014}
\begin{subequations}
\label{eq:cont:PS}
\begin{alignat}{3}
\label{eq:cont:PSa}
{m_i\dot \omega _i}(t) &= - d_i\omega _i(t) +  P_i^* -  {P_i^{net}}(t)+\ v_i(t)\\
\label{eq:cont:PSb}
{\dot P_i^{net}}(t) &= \sum\limits_{j = 1,j \ne i}^N {{E_i}{E_j}\left| {{Y_{ij}}} \right|\left( {{\omega _i} - {\omega _j}} \right)}, i \in \mathcal{N}
\end{alignat}
\end{subequations}
where  ${E_i}>0$ is the nodal voltage, and $Y_{ij}$ is the admittance among buses $i$ and $j$. To achieve synchronization, we consider the distributed control protocol  
\begin{equation}
\label{PIDcontr:PWS}
{v_i}(t) = \frac{1}{m_i}\left(k_i\omega_i(t) + \sigma_P \sum\nolimits_{k = 1}^N {{\mathcal{L}_{P,ij}}{\omega _i}(t)}\right)
\end{equation}
with $k_i\in\amsbb{R}$ being a constant representing a local feedback gain for the $i$th-node, $\sigma_P>0$ and $\Lp_P\in\amsbb{W}$ representing the Laplacian matrix of the proportional layer $\mathscr{G}_{P}$ with link weights $\alpha_{ij}$. 
Now, let $\beta_{ij}:={E_i}{E_j}\left| {{Y_{ij}}} \right|$ be the weights on each edge of the power network in \eqref{eq:cont:PSb} and $\Li_I\in\amsbb{W}$ the associated Laplacian matrix describing the equivalent distributed integral action \eqref{eq:cont:PSb}. Setting $z(t)=-(1/m_i) P_i^{net}(t)$, the problem becomes that of proving convergence in the heterogeneous network given by
\begin{subequations}
\label{eq:exampl:1}
\begin{alignat}{3}
\label{eq:exampl:1a}
\dot {{\boldsymbol\omega }}(t) &= \left( {\mathbf{H} -\sigma_P \Lp_P} \right){\boldsymbol\omega}(t) + {\mathbf{z}}(t) + \mathbf{B}\\
\label{eq:exampl:1b}
\dot {\mathbf{z}}(t) &=  - \mathbf{M}\Li_I\boldsymbol\omega(t)
\end{alignat}
\end{subequations}
where\ ${\boldsymbol\omega}(t):=[{\omega _1}(t), \cdots ,{\omega _N}(t)]$, ${\mathbf{z}}(t):=[{z _1}(t), \cdots ,{z _N}(t)]$ are the stack vectors of frequencies and rescaled electrical powers respectively, $\mathbf{H}:=\mbox{diag}\{k_1-d_1/m_1,\cdots,k_N-d_N/m_N\}$, $\mathbf{M}:=\mbox{diag}\{1/m_1,\cdots,1/m_N\}$ and the vector $\mathbf{B} := \mbox{diag}\{P_1^*/m_1,\cdots,P_N^*/m_N\}$. The closed-loop power system \eqref{eq:exampl:1} has the same structure of the multiplex network $\eqref{eq:DMPI}$ where the input biases $\mathbf{b}_i$ represent the rescaled constant power injections $P_i^*/m_i$ of each node.
\begin{prop}
\label{equil_point:powergrid}
The closed-loop power network \eqref{eq:exampl:1} has a unique equilibrium given by ${\boldsymbol\omega^*}:=\omega_{\infty}\Vone_{N}$, with $\omega_{\infty}:=-{{\sum\nolimits_{i = 1}^N {{P_i^*}} }} / {{\sum\nolimits_{i = 1}^N {( {m_ik_i-d_i} )} }}$ and ${\mathbf{z}^* }:=- ( {\omega_{\infty}\mathbf{H}\Vone_{N} + \mathbf{B} } )$
\end{prop}
\begin{pf*}{Proof.} As done in the proof of Proposition \ref{equil_point}, by setting the left-hand side of \eqref{eq:exampl:1} to zero, one has that $\mathbf{x}^* = a\Vone_N$, $\forall a \in \amsbb{R}$ and ${\mathbf{z}^* } =  -\left( {a\mathbf{P}\Vone + \mathbf{B} } \right)$. Now letting $\mathbf{v}:=[m_1,\cdots,m_N]^T$, by the definition of $\mathbf{z}(t)$ one has that $\mathbf{v}^T\mathbf{z}(t)=0$. Therefore $\mathbf{v}^T\mathbf{z}^*  = 0$ and we obtain $a=  - {\mathbf{v}^T\mathbf{B}}/{\mathbf{v}^T\mathbf{H}{\Vone_N}}=:\omega_{\infty}$
%
%
%
\end{pf*}
%
%
%
%
%
\begin{cor} Under the control dynamics \eqref{PIDcontr:PWS}, the power network \eqref{eq:cont:PS} with $m_i=m,m>0$ $\forall i\in \mathcal{N}$ asymptotically converges to $\omega_{\infty}$ if the following conditions are satisfied
\begin{subequations}
\begin{equation}
\label{eq:c1}
{{\psi _{11}}}=\sum\limits_{i = 1}^N {\left( {{k_i} - \frac{{{d_i}}}{{{m}}}} \right)}  < 0
\end{equation}
\begin{equation}
\label{eq:c2}
\sigma_P {\lambda _2}\left( P \right) > \frac{{\sum\nolimits_{i = 1}^N {{{\left( {{k_i} - \frac{{{d_i}}}{{{m}}}} \right)}^2}} }}{{N\left| {{\psi _{11}}} \right|}} + \mathop {\max }\limits_i \left\{ {{k_i} - \frac{{{d_i}}}{{{m}}}} \right\}
\end{equation}
\end{subequations}
\label{Coll:PI}
\end{cor}
\begin{pf*}{Proof.} Note that \eqref{eq:exampl:1} can be seen as a group of $N$ first order heterogeneous agents controlled by a multiplex PI strategy. Specifically, letting $A_i={k_i} - {d_i}/{m}$, the dynamics of each node can be written as
\[\begin{array}{l}
{{\dot \omega }_i}(t) = A_i{\omega _i}(t) + {b_i} - \sigma_P \sum\nolimits_{j = 1}^N {{\mathcal{L}_{P,ij}}{\omega _j}(t)}  + {z_i}(t)\\
{{\dot z}_i}(t) =  - (1/m)\sum\nolimits_{j = 1}^N {{\mathcal{L}_{I,ij}}{\omega _j}(t)} 
\end{array}\]
Therefore, using Theorem \eqref{Th:I:PI} with $\sigma=0$, and $\sigma_I = (1/m)$ completes the proof.
\end{pf*}
\subsection{Illustrative example}
\begin{figure}[tbp]
\centering {
\subfigure[]
{\label{fig:PGa}
{\includegraphics[scale=0.134]{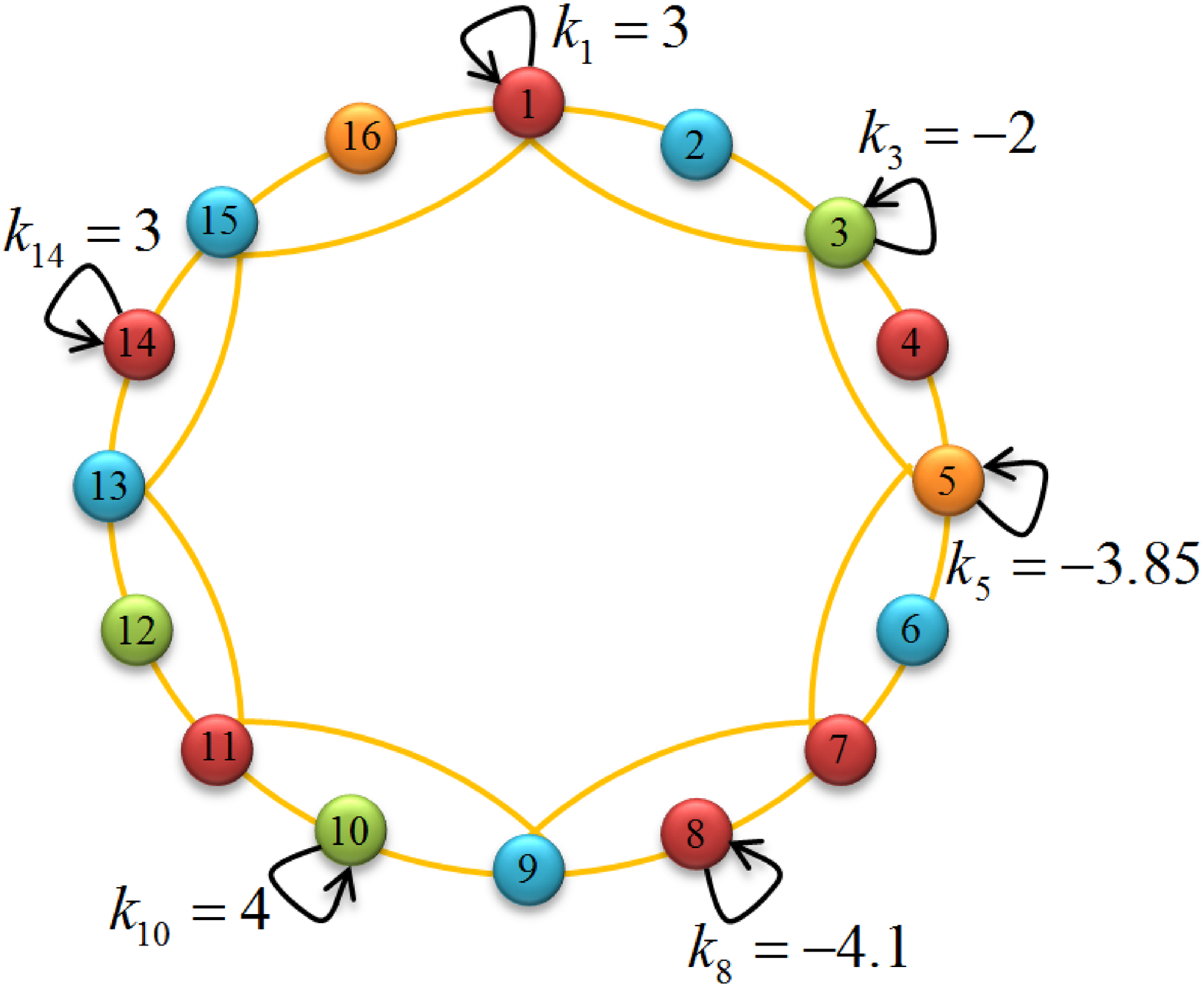}}}
\subfigure[]
{\label{fig:PGb}
{\includegraphics[scale=0.134]{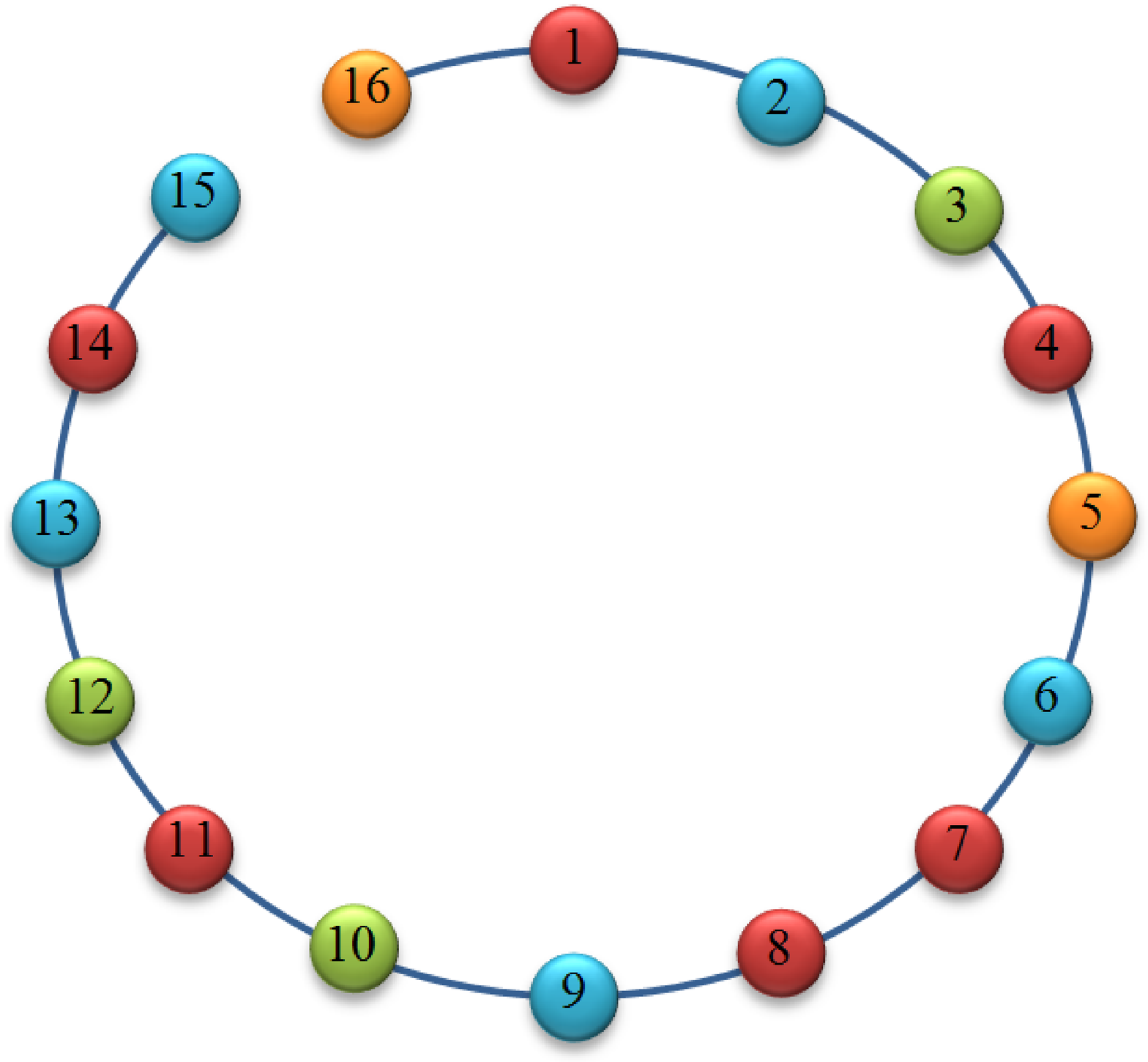}}}
\subfigure[]
{\label{fig:PGc}
{\includegraphics[scale=0.5]{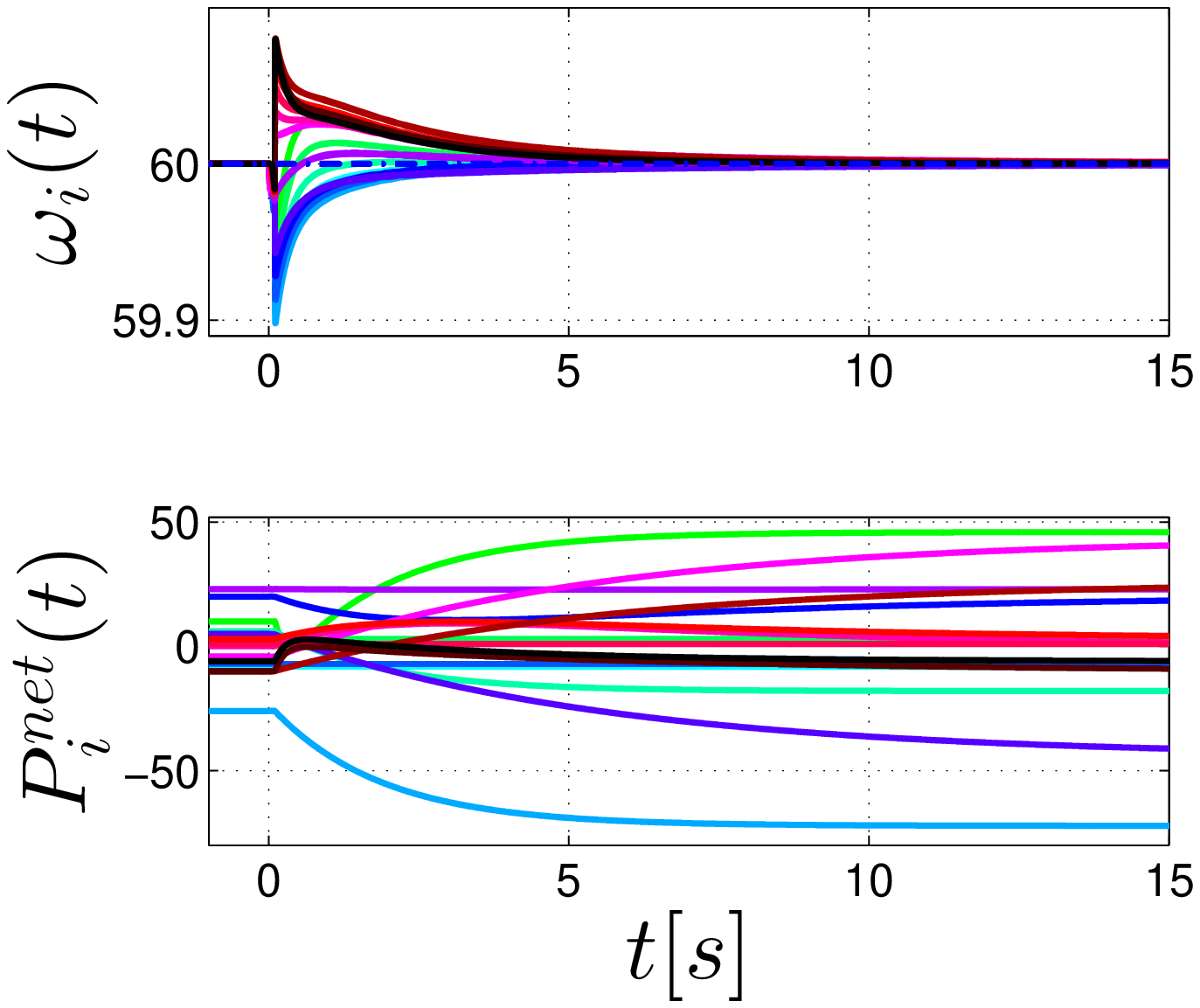}}}
}
\caption{(a),(b) Network architectures representing the integral and proportional control layers respectively. The gains of the proportional layer are set as $\alpha_{ij}=200$. (c) Evolution of the power network. The blue dash-dot line represent the convergence value $\omega_{\infty}$.}
\label{fig:F_E2} 
\end{figure}

%
%
As an illustration, consider the power network shown in Fig. \ref{fig:PGa}. For the sake of simplicity, and without loss of generality, we consider all line admittances and nodal voltages to be $Y_{ij}=0.0001$ and $ E_i=2kV$ $\forall i,j\in \mathcal{N}$ respectively. Moreover, we assume $m=0.2$ and four different values of damping, that is $d_i=0.5$, for $i\in\{1,4,7,8,11,14\}$, $d_i=0.45$, $i\in\{2,6,9,13,15\}$, $d_i=0.40$, $i\in\{3,10,12\}$, while $d_i=0.6$, $i\in\{5,16\}$. Furthermore the vector containing the nominal power injections (expressed in MW) for each node is given by $\mathbf{P}^{*} = [40, 30, 30, 22, 10, 20, 50, 35, 50, 20, 30, 25, 30, 20, 17, 30]$.
Following the approach in \cite{Andreasson2014}, we assume that the network has been operating in these nominal conditions for $t<0$ [see Fig.  \ref{fig:PGc}]. As the power network \eqref{eq:exampl:1} has a natural integral controller which encode the phase angles $\delta_i(t)$, consensus is expected on a value dependent on the network parameters and the nominal power injections. Such a value can be easily computed from Proposition \ref{equil_point:powergrid} by setting all $k_i =0$ yielding $\omega_{\infty}=60Hz$.

Now, consider the scenario where, at $t=0s$ the nominal power injections are decreased by 600kW from the nominal value at buses $4$, $8$ and $10$ and consequently, the frequencies of all generators decrease as well. 
To compensate those disturbances, we use local feedback controllers on a fraction of nodes together with a distributed proportional action to manipulate and stabilize the desired convergence value. Specifically, we introduce feedback controllers with appropriate gains at nodes $1$, $3$, $5$, $8$, $10$ and $14$ [denoted by self feedback loops indicated in black in Fig. \ref{fig:PGa})] in order to shift $\omega_{\infty}$ to the desired value $\omega_{\infty}=60$. To address the stability of such consensus equilibrium, we use Corollary \ref{Coll:PI}. 
Firstly we find that ${{\psi _{11}}}:=-2.3875$ and condition \eqref{eq:c1} is fulfilled. Secondly, we have that $\mathop {\max }\nolimits_i \left\{ {{k_i} - {{{d_i}}}/{{{m}}}} \right\}=2$ and therefore the power network reaches admissible consensus if $\sigma_P {\lambda _2}\left( \Lp_P \right) > 6.3991$. Choosing a simple path graph for the proportional control layer as shown in Fig. \ref{fig:PGb} yields $\sigma_P>0.8326$ to guarantee convergence. Heuristically, we found that setting $\sigma_P$ = 55 also ensures that the maximum frequency overshoot during transient is less than $100mHz$ (necessary to avoid unwanted damage to the grid).
The behaviour of the closed-loop power network is shown in Fig. \ref{fig:PGc} where the distributed controller is switched on at $t=0.1$. As expected we observe the power network to quickly regain stability onto the desired target frequency.
%
%
%
%
%
%
%
%
\section{Conclusions}
We have proposed a novel approach for controlling networks of heterogeneous nodes with generic $n$-dimensional linear dynamics in the presence of constant biases (disturbances).
In particular, we discussed the use of different control layers, each with its own topology, deploying proportional and integral actions across the network. We proved convergence of the strategy and derived conditions to select the control gains as a function of the open loop and control network structures and the node dynamics. We showed the effectiveness of the proposed strategy via numerical simulations on two representative examples.

Several open problems are left for further study. First and foremost the effect of varying the structure of the network control layers should be studied in more detail as preliminary results show the performance of the network evolution towards consensus can be affected by such variations.  We wish to emphasize that more sophisticated approaches can be developed by considering other linear or nonlinear control actions rather than the simpler proportional and integral actions considered in this paper. For example a robustifying distributed action could be designed by considering an extra network control layer of variable structure controllers. This is currently under investigation and will be presented elsewhere.



\bibliographystyle{plain}        
\bibliography{Main}            

\begin{thebibliography}{10}

\bibitem{Andreasson2014}
M.~Andreasson, D.V. Dimarogonas, H.~Sandberg, and K.H. Johansson.
\newblock Distributed control of networked dynamical systems: Static feedback,
  integral action and consensus.
\newblock {\em IEEE Transactions on Automatic Control}, 2014.

\bibitem{Bai2010}
H.~Bai, R.A. Freeman, and K.M. Lynch.
\newblock Robust dynamic average consensus of time-varying inputs.
\newblock In {\em 49th IEEE Conference on Decision and Control (CDC)}, pages
  3104--3109, 2010.

\bibitem{Bai2014}
H.~Bai and S.Y. Shafi.
\newblock Output synchronization of nonlinear systems under input disturbances.
\newblock In {\em 21st International Symposium on Mathematical Theory of
  Networks and Systems}, pages 506--512, 2014.

\bibitem{Bernstein2009}
D.S. Bernstein.
\newblock {\em Matrix Mathematics: Theory, Facts, and Formulas (Second
  Edition)}.
\newblock Princeton University Press, 2009.

\bibitem{Bidram2014}
A.~Bidram, F.L. Lewis, and A.~Davoudi.
\newblock Distributed control systems for small-scale power networks: Using
  multiagent cooperative control theory.
\newblock {\em IEEE Control Systems Magazine}, 34(6):56--77, 2014.

\bibitem{Bondy2008}
J.A. Bondy and U.S.R. Murty.
\newblock {\em Graph Theory}.
\newblock Springer, 2008.

\bibitem{Burbano2014ISCAS}
D.A. Burbano~L. and M.~di~Bernardo.
\newblock Consensus and synchronization of complex networks via
  proportional-integral coupling.
\newblock In {\em IEEE International Symposium on Circuits and Systems
  (ISCAS)}, pages 1796--1799, 2014.

\bibitem{BurbanoLombana2015}
D.A. Burbano~L. and M.~di~Bernardo.
\newblock Distributed {PID} control for consensus of homogeneous and
  heterogeneous networks.
\newblock {\em IEEE Transactions on Control of Network Systems}, 2(2):154--163,
  2015.

\bibitem{Burger2014}
M.~B\"{u}rger and C.~De~Persis.
\newblock Further result about dynamic coupling for nonlinear output agreement.
\newblock In {\em IEEE 53rd Annual Conference on Decision and Control (CDC)},
  pages 1353--1358, 2014.

\bibitem{CarliCSZ2008}
R.~Carli, A.~Chiuso, L.~Schenato, and S.~Zampieri.
\newblock A {PI} consensus controller for networked clocks synchronization.
\newblock In {\em 17th IFAC World Congress}, volume~17, pages 10289--10294,
  2008.

\bibitem{CHEN2013}
G.~Chen.
\newblock Problems and challenges in control theory under complex dynamical
  network environments.
\newblock {\em Acta Automatica Sinica}, 39(4):312 -- 321, 2013.

\bibitem{Cornelius2013}
S.P. Cornelius, W.L. Kath, and A.E. Motter.
\newblock Realistic control of network dynamics.
\newblock {\em Nature Communications}, 4(1942), 2013.

\bibitem{DeLellis2015}
P.~DeLellis, M.~di~Bernardo, and D.~Liuzza.
\newblock Convergence and synchronization in heterogeneous networks of smooth
  and piecewise smooth systems.
\newblock {\em Automatica}, 56(6):1--11, 2015.

\bibitem{Freeman2006}
R.A. Freeman, Peng Yang, and K.M. Lynch.
\newblock Stability and convergence properties of dynamic average consensus
  estimators.
\newblock In {\em Proceedings of 45th IEEE Conference on Decision and Control},
  pages 338 --343, 2006.

\bibitem{Garulli2011}
A.~Garulli and A.~Giannitrapani.
\newblock Analysis of consensus protocols with bounded measurement errors.
\newblock {\em Systems and Control Letters}, 60(1):44 -- 52, 2011.

\bibitem{Hill2006}
D.J. Hill and G.~Chen.
\newblock Power systems as dynamic networks.
\newblock In {\em Proceedings of International Symposium on Circuits and
  Systems {ISCAS}}, pages 722--725, 2006.

\bibitem{RAH_CRJ_1987}
A.R. Horn and R.C. Johnson.
\newblock {\em Matrix Analysis}.
\newblock Cambridge Univ. Press, 1987.

\bibitem{Kenji2011}
F.~Kenji and H.~Yuko.
\newblock On convergence rate of distributed consensus for homogeneous graphs.
\newblock {\em Proceedings of the 18th IFAC World Congress}, 18:10032--10037,
  2011.

\bibitem{Li2010}
Z.~Li, Z.~Duan, G.~Chen, and L.~Huang.
\newblock Consensus of multiagent systems and synchronization of complex
  networks: A unified viewpoint.
\newblock {\em IEEE Transactions on Circuits and Systems I}, 57(1):213--224,
  2010.

\bibitem{Liu2011}
Shuai Liu, Lihua Xie, and Huanshui Zhang.
\newblock Distributed consensus for multi-agent systems with delays and noises
  in transmission channels.
\newblock {\em Automatica}, 47(5):920 -- 934, 2011.

\bibitem{LiuS2011}
Y.Y. Liu and Barabási~A.L. Slotine, J.J.
\newblock Controllability of complex networks.
\newblock {\em Nature}, 473:167--173, 2011.

\bibitem{Lu2006214}
W.~Lu and T.~Chen.
\newblock New approach to synchronization analysis of linearly coupled ordinary
  differential systems.
\newblock {\em Physica D: Nonlinear Phenomena}, 213(2):214 -- 230, 2006.

\bibitem{Lunze2012}
J.~Lunze.
\newblock Synchronization of heterogeneous agents.
\newblock {\em IEEE Transactions on Automatic Control}, 57(11):2885--2890,
  2012.

\bibitem{Meng2014}
Deyuan Meng and K.L. Moore.
\newblock Studies on resilient control through multiagent consensus networks
  subject to disturbances.
\newblock {\em IEEE Transactions on Cybernetics}, 44(11):2050--2064, 2014.

\bibitem{Motter2013power}
A.E. Motter, S.A Myers, M.~Anghel, and T.~Nishikawa.
\newblock Spontaneous synchrony in power-grid networks.
\newblock {\em Nature Physics}, 9:191–197, 2013.

\bibitem{Mucha2010}
P.J. Mucha, T.~Richardson, K.~Macon, M.~A. Porter, and J.P. Onnela.
\newblock Community structure in time-dependent, multiscale, and multiplex
  networks.
\newblock {\em Science}, 328(5980):876--878, 2010.

\bibitem{OlfatFM2007}
R.~Olfati-Saber, J.A. Fax, and R.M. Murray.
\newblock Consensus and cooperation in networked multi-agent systems.
\newblock {\em Proceedings of the IEEE}, 95(1):215 --233, 2007.

\bibitem{Poole1974}
G.~Poole and T.~Boullion.
\newblock A survey on {M}-matrices.
\newblock {\em SIAM Review}, 16(4):419--427, 1974.

\bibitem{Ren2007c}
W.~Ren and W.~Beard.
\newblock {\em Distributed Consensus in Multi-vehicle Cooperative Control:
  Theory and Applications}.
\newblock Springer-Verlag, 2007.

\bibitem{ReCa:11}
W.~Ren and Y.~Cao.
\newblock {\em Distributed Coordination of Multi-agent Networks}.
\newblock Springer-Verlag, 2011.

\bibitem{Ren2007HO}
W.~Ren, K.~L. Moore, and Y.~Chen.
\newblock High-order and model reference consensus algorithms in cooperative
  control of multivehicle systems.
\newblock {\em Dynamic Systems, Measurement, and Control}, 129:678--688, 2007.

\bibitem{Sarlette2012}
A.~Sarlette, J.~Dai, Y.~Phulpin, and D.~Ernst.
\newblock Cooperative frequency control with a multi-terminal high-voltage {DC}
  network.
\newblock {\em Automatica}, 48(12):3128 -- 3134, 2012.

\bibitem{ScardS2009}
L.~Scardovi and R.~Sepulchre.
\newblock Synchronization in networks of identical linear systems.
\newblock {\em Automatica}, 45(11):2557 -- 2562, 2009.

\bibitem{Seyboth2015}
G.~Seyboth and F.~Allg\"{o}wer.
\newblock Output synchronization of linear multi-agent systems under constant
  disturbances via distributed integral action.
\newblock In {\em American Control Conference (ACC). Chicago, IL, USA.}, pages
  62--67, 2015.

\bibitem{Seyboth2012}
G.~S. Seyboth, D.~V. Dimarogonas, K.~H. Johansson, and F.~Allg\"{o}wer.
\newblock Static diffusive couplings in heterogeneous linear networks.
\newblock In {\em 3rd IFAC Workshop on Distributed Estimation and Control in
  Networked Systems}, pages 258--263, 2012.

\bibitem{Simpson-Porco2013}
J.W. Simpson-P., F.~D\"{o}rfler, and F.~Bullo.
\newblock Synchronization and power sharing for droop-controlled inverters in
  islanded microgrids.
\newblock {\em Automatica}, 49(9):2603 -- 2611, 2013.

\bibitem{Wang2014}
Z.~Wang, J.~Xu, and H.~Zhang.
\newblock Consensusability of multi-agent systems with time-varying
  communication delay.
\newblock {\em Systems \& Control Letters}, 65:37 -- 42, 2014.

\bibitem{ZhongOct}
Z.~Wei-Song, L.~Guo-Ping, and C.~Thomas.
\newblock Global bounded consensus of multiagent systems with nonidentical
  nodes and time delays.
\newblock {\em IEEE Transactions on Systems, Man, and Cybernetics, Part B:
  Cybernetics}, 42(5):1480--1488, 2012.

\bibitem{Wieland2009}
P.~Wieland and F.~Allg\"{o}wer.
\newblock An internal model principle for consensus in heterogeneous linear
  multi-agent systems.
\newblock In {\em 1st IFAC Workshop on Estimation and Control of Networked
  Systems}, volume~1, pages 7--12, 2009.

\bibitem{Wieland2011}
P.~Wieland, R.~Sepulchre, and Allg\"{o}wer F.
\newblock An internal model principle is necessary and sufficient for linear
  output synchronization.
\newblock {\em Automatica}, 47(5):1068 -- 1074, 2011.

\bibitem{Wieland2013}
P.~Wieland, Jingbo Wu, and F.~All\"{o}gwer.
\newblock On synchronous steady states and internal models of diffusively
  coupled systems.
\newblock {\em IEEE Transactions on Automatic Control}, 58(10):2591--2602,
  2013.

\bibitem{Zhang2014}
Z.~Xuan and A.~Papachristodoulou.
\newblock A distributed {PID} controller for network congestion control
  problems.
\newblock In {\em American Control Conference (ACC)}, pages 5453--5458, 2014.

\end{thebibliography}


\appendix
\section{Proof of Lemma \ref{lemm:simmetric_L}}
\label{Appendix_II}
As the Laplacian matrix is symmetric (the graph is undirected), according to Schur's lemma, there exists an orthogonal matrix, say $ {\mathbf{V}}$ such that ${\WideLaplacian}=  {\mathbf{V}} \mathbf{\Lambda}  {\mathbf{V}}^{-1}$ where the eigenvectors of ${\WideLaplacian}$ are column vectors of $\mathbf{V}$ (or equivalently row vectors of $\mathbf{V}^{-1}$). The eigenvector associated with the null eigenvalue of  ${\WideLaplacian}$ is given by $\mathbf{v}_1=[1/\sqrt{N},\cdots,1/\sqrt{N}]$. Then, rewriting $\mathbf{V}$ in block form one has that
\[\mathbf{V} = \left[ {\begin{array}{*{20}{c}}
{{V_{11}}}&{{\mathbf{V}_{12}}}\\
{{\mathbf{V}_{21}}}&{{\mathbf{V}_{22}}}
\end{array}} \right]\]
where $V_{11}=1/\sqrt{N}$ and $\mathbf{V}_{21}=(1/\sqrt{N})\Vone_{N-1}^T$, $\mathbf{V}_{12}\in {\amsbb{R}^{1 \times (N - 1)}}$, and $\mathbf{V}_{22}\in \amsbb{R}^{{(N - 1) \times (N - 1)}}$. Then, some straightforward algebra yields,
\[{\WideLaplacian} = \left[ {\begin{array}{*{20}{c}}
1&{\sqrt{N}{\mathbf{V}_{12}}}\\
{{\Vone_{N - 1}}}&{\sqrt{N}{\mathbf{V}_{22}}}
\end{array}} \right]\mathbf{\Lambda} \left[ {\begin{array}{*{20}{c}}
{1/N}&{(1/N)\Vone_{N - 1}^T}\\
{(1/\sqrt N )\mathbf{V}_{12}^T}&{(1/\sqrt N )\mathbf{V}_{22}^T}
\end{array}} \right]\]
Thus, setting $r_{11} = {1}/{N}$, $\mathbf{R}_{12}= {1}/{N}\Vone_{N - 1}^T$, $\mathbf{R}_{21}={(1/\sqrt N )\mathbf{V}_{12}^T}$ and $\mathbf{R}_{22} = {(1/\sqrt N )\mathbf{V}_{22}^T}$ we obtain \eqref{block:decompo}.
Also, since $\mathbf{V}^{-1} \mathbf{V}=\mathbf{R}^{-1} \mathbf{R}=\mathbf{I}_N$, the blocks in the definition of $\mathbf{R}$ and $\mathbf{R}^{-1}$ must fulfill conditions $\eqref{prop:U:1}-\eqref{prop:U:5}$. Moreover, ${\specnorm{ \mathbf{V}^{-1} } }=\sqrt{\lambda_{max}( (\mathbf{V}^{-1})^T {\mathbf{V}}^{-1})}$. Also, as $ {\mathbf{V}}^{-1}= {\mathbf{V}}^{T}$ and $\mathbf{V}\mathbf{V}^T=\mathbf{I}_{N}$ one has {$\mathbf{R}^T=N\mathbf{R}^{-1}$; therefore, $\specnorm{{\mathbf{R}^{ - 1}}} = {1}/{{\sqrt N }}$ and $\specnorm{{\mathbf{R}_{22}}} \le \specnorm{{\mathbf{R}^{ - 1}}}$ that together with \eqref{kron:rel:c} yields \eqref{prop:U:5norm} }.
%
%
%
%
\section{Proof of Lemma \ref{prop:dif_lap}}
\label{Appendix_I}
Multiplying both sides of $\WideLaplacian_2=\mathbf{U}\mathbf{\Lambda_2}\mathbf{U}^{-1}$ by $\mathbf{R}^{-1}$ and $\mathbf{R}$, yields $\mathbf{R}^{-1}\WideLaplacian_2\mathbf{R}=\mathbf{R}^{-1}\mathbf{U}\mathbf{\Lambda_2}\mathbf{U}^{-1}\mathbf{R}$.
Now using the block form of $\mathbf{R}$ and $\mathbf{U}$ as shown in Lemma \ref{lemm:simmetric_L} one has that $\mathbf{R}^{-1}\WideLaplacian_2\mathbf{R}$ is given by
\[
\begin{split}
\left[ {\begin{array}{*{20}{c}}
{{r_{11}}}&{{\mathbf{R}_{12}}}\\
{{\mathbf{R}_{21}}}&{{\mathbf{R}_{22}}}
\end{array}} \right] \cdot \left[ {\begin{array}{*{20}{c}}
1&{N\mathbf{U}_{21}^T}\\
{{\Vone_{N - 1}}}&{N\mathbf{U}_{22}^T}
\end{array}} \right]  \cdot \left[ {\begin{array}{*{20}{c}}
0&{{\Vzero_{1 \times (N - 1)}}}\\
{{\Vzero_{(N - 1) \times 1}}}&{{{\bar {\mathbf{\Lambda}} }_2}}
\end{array}} \right] \\
 \cdot \left[ {\begin{array}{*{20}{c}}
{{u_{11}}}&{{\mathbf{U}_{12}}}\\
{{\mathbf{U}_{21}}}&{{\mathbf{U}_{22}}}
\end{array}} \right] \cdot \left[ {\begin{array}{*{20}{c}}
1&{N\mathbf{R}_{21}^T}\\
{{\Vone_{N - 1}}}&{N\mathbf{R}_{22}^T}
\end{array}} \right]
\end{split}
\]
where $\mathbf{\bar{\Lambda}}_2=\mbox{diag}\left\{\lambda_2(\WideLaplacian_2),\cdots,\lambda_N(\WideLaplacian_2)\right\}$. By definition $u_{11}=r_{11}$ and $\mathbf{U}_{12}=\mathbf{R}_{12}$ (see \eqref{eq:blockdef}), and by some matrix manipulation we obtain
\begin{equation}
\label{proof:lem:dif:lap}
\begin{split}
\mathbf{R}^{-1}\WideLaplacian_2\mathbf{R} &= \left[ {\begin{array}{*{20}{c}}
{{r_{11}} + {\mathbf{R}_{12}}{\Vone_{N - 1}}} & {N({u_{11}}\mathbf{U}_{21}^T + {\mathbf{U}_{12}}\mathbf{U}_{22}^T)}\\
{{\mathbf{R}_{21}} + {\mathbf{R}_{22}}{\Vone_{N - 1}}}&{N({\mathbf{R}_{21}}\mathbf{U}_{21}^T + {\mathbf{R}_{22}}\mathbf{U}_{22}^T)}
\end{array}} \right] \\ 
& \quad \left[ {\begin{array}{*{20}{c}}
0&{{\Vzero_{1 \times (N - 1)}}}\\
{{\Vzero_{(N - 1) \times 1}}}&{{{\bar {\mathbf{\Lambda}} }_2}}
\end{array}} \right]\\
& \quad \left[ {\begin{array}{*{20}{c}}
{{u_{11}} + {\mathbf{U}_{12}}{\Vone_{N - 1}}}&{N({u_{11}}\mathbf{R}_{21}^T + {\mathbf{R}_{12}}\mathbf{R}_{22}^T)}\\
{{\mathbf{U}_{21}} + {\mathbf{U}_{22}}{\Vone_{N - 1}}}&{N({\mathbf{U}_{21}}\mathbf{R}_{21}^T + {\mathbf{U}_{22}}\mathbf{R}_{22}^T)}
\end{array}} \right] 
\end{split}
\end{equation}
We next simplify each block of all matrices. Then, from \eqref{prop:U:1} we have that ${{r_{11}} + {\mathbf{R}_{12}}{\Vone_{N - 1}}}={{u_{11}} + {\mathbf{U}_{12}}{\Vone_{N - 1}}}=1$. While, from \eqref{prop:U:2} ${{\mathbf{R}_{21}} + {\mathbf{R}_{22}}{\Vone_{N - 1}}}={{\mathbf{U}_{21}} + {\mathbf{U}_{22}}{\Vone_{N - 1}}}=\Vzero$ and using \eqref{prop:U:4} ${N({u_{11}}\mathbf{U}_{21}^T + {\mathbf{U}_{12}}\mathbf{U}_{22}^T)}=N({r_{11}}\mathbf{R}_{21}^T + $ ${\mathbf{R}_{12}}\mathbf{R}_{22}^T)=\Vzero$.
Note also that ${{\mathbf{R}_{21}}=-{\mathbf{R}_{22}}{\Vone_{N - 1}}}$ and ${{\mathbf{U}_{21}} =- {\mathbf{U}_{22}}{\Vone_{N - 1}}}$. Thus, the blocks
\begin{equation}
\label{xi_a}
\begin{split}
\mathbf{T}_1 &:= N({\mathbf{R}_{21}}\mathbf{U}_{21}^T + {\mathbf{R}_{22}}\mathbf{U}_{22}^T)\\
             &= N\mathbf{R}_{22}(\Vone_{N-1}\Vone_{N-1}^T+\mathbf{I}_{N-1})\mathbf{U}_{22}^T
\end{split}
\end{equation}
and,
\begin{equation}
\label{xi_b}
\begin{split}
\mathbf{T}_2 &:= N({\mathbf{U}_{21}}\mathbf{R}_{21}^T + {\mathbf{U}_{22}}\mathbf{R}_{22}^T)\\
              &= N\mathbf{U}_{22}(\Vone_{N-1}\Vone_{N-1}^T+\mathbf{I}_{N-1})\mathbf{R}_{22}^T
\end{split}
\end{equation}
Consequently, we have $\mathbf{T}_1=\mathbf{T}_2^T$ and letting $\mathbf{T}=\mathbf{T}_1$, the Kronecker product $(\mathbf{R}^{-1}\WideLaplacian_2\mathbf{R} \otimes \mathbf{I}_n)$ yields \eqref{lem:dif:lap}. Finally, to prove that $\mathbf{T}{{{\bar {\mathbf{\Lambda}} }_2}}\mathbf{T}^T$ is a symmetric matrix we have to show that $\mathbf{T}^T$ is an orthonormal matrix. Then, from \eqref{xi_a} and from the fact that $\mathbf{R}_{22}$ is an invertible (full rank) matrix \cite{BurbanoLombana2015} one has %
$\mathbf{T}^{-1}={1}/{N}(\mathbf{U}_{22}^T)^{-1}(\Vone_{N-1}\Vone_{N-1}^T+\mathbf{I}_{N-1})^{-1}\mathbf{R}_{22}^{-1}$
and using property \eqref{prop:U:7inv} we obtain $\mathbf{T}^{-1} = N\mathbf{U}_{22}(\Vone_{N-1}\Vone_{N-1}^T+\mathbf{I}_{N-1})\mathbf{R}_{22}^T = \mathbf{T}^T$ which completes the proof.
\section{Derivation of $\mathbf{\Psi}$}
\label{Appendix_III}
Using the block decomposition of $\mathbf{R}$ as done in Appendix \ref{Appendix_I}, we have
\[
\begin{split}
\mathbf{\Psi}&:=\left[ {\begin{array}{*{20}{c}}
{{\mathbf{\Psi} _{11}}}&{{\mathbf{\Psi} _{12}}}\\
{{\mathbf{\Psi} _{21}}}&{{\mathbf{\Psi} _{22}}}
\end{array}} \right] = \left[ {\begin{array}{*{20}{c}}
{{r_{11}}{\mathbf{I}_n}}&{({\mathbf{R}_{12}} \otimes {\mathbf{I}_n})}\\
{({\mathbf{R}_{21}} \otimes {\mathbf{I}_n})}&{({\mathbf{R}_{22}} \otimes {\mathbf{I}_n})}
\end{array}} \right] \cdot \\
 &\quad \left[ {\begin{array}{*{20}{c}}
{\mathbf{A}_1}&{{\Vzero}}\\
{{\Vzero}}&{\bar {\mathbf{A}}}
\end{array}} \right]
 \cdot \left[ {\begin{array}{*{20}{c}}
{{\mathbf{I}_n}}&{N(\mathbf{R}_{21}^T \otimes {\mathbf{I}_n})}\\
{({\Vone_{N - 1}} \otimes {\mathbf{I}_n})}&{N(\mathbf{R}_{22}^T \otimes {\mathbf{I}_n})}
\end{array}} \right]
\end{split}
\]
%
%
%
%
where $\mathbf{\Psi} _{11} = {r_{11}}\mathbf{A}_1 + ({\mathbf{R}_{12}} \otimes {\mathbf{I}_n})\bar {\mathbf{A}}({\Vone_{N - 1}} \otimes {\mathbf{I}_n})$, $\mathbf{\Psi}_{12} = N\left({ {{r_{11}}\mathbf{A}_1(\mathbf{R}_{21}^T \otimes {\mathbf{I}_n}) + (\mathbf{R}_{12} \otimes {\mathbf{I}_n})\bar {\mathbf{A}}(\mathbf{R}_{22}^T \otimes {\mathbf{I}_n})} }\right)$, $\mathbf{\Psi}_{21} = ({\mathbf{R}_{21}} \otimes {\mathbf{I}_n})\mathbf{A}_1 + ({\mathbf{R}_{22}} \otimes {\mathbf{I}_n})\bar {\mathbf{A}}({\Vone_{N - 1}} \otimes {\mathbf{I}_n})$, and $\mathbf{\Psi}_{22}= N ({\mathbf{R}_{21}} \otimes {\mathbf{I}_n})\mathbf{A}_1(\mathbf{R}_{21}^T \otimes {\mathbf{I}_n}) + N({\mathbf{R}_{22}} \otimes {\mathbf{I}_n})\bar {\mathbf{A}}(\mathbf{R}_{22}^T \otimes {\mathbf{I}_n})$

Now, by some algebraic manipulations we can simplify each block of $\mathbf{\Psi}$. Then, by definition $r_{11}=1/N$ and $\mathbf{R}_{12}=(1/N)\Vone_{N-1}^T$ and $\mathbf{\Psi} _{11}=(1/N)(\mathbf{A}_1 + ({\Vone_{N - 1}^T} \otimes {\mathbf{I}_n}) \bar{\mathbf{A}} ({\Vone_{N - 1}} \otimes {\mathbf{I}_n})  )$ which is clearly \eqref{eq:PSI:11}. For the second block we can add and subtract $N\mathbf{A}_1(\mathbf{R}_{12}\mathbf{R}_{22}^T\otimes\mathbf{I}_n)$ where $\mathbf{R}_{12}=(1/N)\Vone_{N - 1}^T$. Hence, using \eqref{prop:U:4} one has
$$\mathbf{\Psi} _{12}=\underbrace{\left(({\Vone_{N - 1}^T} \otimes {\mathbf{I}_n})\bar{\mathbf{A}}- \mathbf{A}_1({\Vone_{N - 1}^T} \otimes {\mathbf{I}_n})\right) }_{\mathbf{P}_1}(\mathbf{R}_{22}^T\otimes\mathbf{I}_n)$$
note that the matrix $\mathbf{P}_1$ can be recast as $\mathbf{P}_1 = [{\mathbf{A}_2}\,\,{\mathbf{A}_3}\,\, \cdots {\mathbf{A}_{N - 1}}] - [{\mathbf{A}_1}\,\,{\mathbf{A}_1}\, \cdots {\mathbf{A}_1}] = [{\mathbf{A}_2} - {\mathbf{A}_1} \cdots {\mathbf{A}_{N}} - {\mathbf{A}_1}]$ and then \eqref{eq:PSI:12} is obtained. Then, following a similar procedure as done before but for $\mathbf{\Psi} _{21}$ adding and subtracting $(\mathbf{R}_{22}\Vone_{N-1}\otimes\mathbf{I}_n)\mathbf{A}_1$, and using property \eqref{prop:U:2} we obtain 
$$\mathbf{\Psi} _{21}=(\mathbf{R}_{22}\otimes\mathbf{I}_n) \underbrace{ \left(\bar{\mathbf{A}}({\Vone_{N - 1}} \otimes {\mathbf{I}_n})  - ({\Vone_{N - 1}} \otimes {\mathbf{I}_n})\mathbf{A}_1 \right) } _{\mathbf{P}_2}$$
in this case $\mathbf{P}_2$ can be rewritten as $\mathbf{P}_2=[{{\mathbf{A}_2^T} - {\mathbf{A}_1}}^T, \cdots,$ ${{\mathbf{A}_N^T} - {\mathbf{A}_1}}^T]^T$.
Finally, from properties \eqref{prop:U:2} and \eqref{prop:U:4} we can express $({\mathbf{R}_{21}} \otimes {\mathbf{I}_n}) =- ({\mathbf{R}_{22}}{\Vone_{N - 1}} \otimes {\mathbf{I}_n})$ and $(\mathbf{R}_{21}^T \otimes {\mathbf{I}_n}) =- (1/{r_{11}})({\mathbf{R}_{12}}\mathbf{R}_{22}^T \otimes {\mathbf{I}_n})$ and the last block reads $\mathbf{\Psi}_{22} := N({\mathbf{R}_{22}} \otimes {\mathbf{I}_n})\widetilde{\mathbf{A}}_1({\mathbf{R}_{22}}^T \otimes {\mathbf{I}_n})+ N({\mathbf{R}_{22}} \otimes {\mathbf{I}_n})\bar {\mathbf{A}}({\mathbf{R}_{22}}^T \otimes {\mathbf{I}_n})$, where $\widetilde{\mathbf{A}}_1:=(\Vone_{N - 1} \otimes {\mathbf{I}_n})\mathbf{A}_1(\Vone_{N - 1}^T \otimes {\mathbf{I}_n})$. Note that $\widetilde{\mathbf{A}}_1$ can also be written as $\widetilde{\mathbf{A}}_1=(\Vone_{N - 1}\Vone_{N - 1}^T  \otimes {\mathbf{A}_1})$ and by grouping common terms we obtain \eqref{eq:PSI:22}.

\end{document}